\documentclass[11pt]{article}
\newif\ifsubmission
\submissionfalse
\ifsubmission\usepackage[review]{acl}\else\usepackage{acl}\fi
\ifsubmission\else
\hypersetup{
  pdftitle={FLASH-MAXSIM: IO-Aware Fused Kernels for Late-Interaction Retrieval},
  pdfauthor={Roi Pony, Daniel Ezer, Adi Raz Goldfarb, Idan Friedman, Oshri Naparstek, Udi Barzelay},
  pdfsubject={EMNLP 2026 Industry Track},
  pdfkeywords={late-interaction retrieval, MaxSim, GPU kernels, IO-aware, ColBERT, ColPali}
}
\fi
\usepackage{times}
\usepackage{latexsym}
\usepackage[T1]{fontenc}
\usepackage[utf8]{inputenc}
\usepackage{microtype}
\usepackage{inconsolata}
\usepackage{graphicx}
\usepackage{amsmath,amssymb}
\usepackage{booktabs}
\usepackage{multirow}
\usepackage{algorithm}
\usepackage{algpseudocode}
\usepackage{float}            
\usepackage{tikz}             
\usetikzlibrary{arrows.meta, positioning}
\usepackage{listings}
\usepackage{xcolor}
\usepackage{placeins}
\usepackage{amsthm}
\newtheorem{theorem}{Theorem}
\newtheorem{proposition}{Proposition}

\newcommand{\repourl}{\ifsubmission \url{https://anonymous.4open.science/r/flash-maxsim-0483}\else\url{https://github.com/roipony/flash-maxsim}\fi}
\newcommand{\repofn}{\footnote{\label{fn:repo}Code, benchmark scripts, and the raw result JSONs behind every numbered table: \repourl{}}}

\newcommand{\maxsim}{\textsc{MaxSim}}
\newcommand{\flash}{\textsc{Flash-MaxSim}}
\newcommand{\flashshort}{\textsc{FM}}

\title{\flash{}: IO-Aware Fused Kernels for Late-Interaction Retrieval}

\ifsubmission
  \author{Anonymous authors\\Affiliation withheld for double-blind review}
\else
  \author{Roi Pony\thanks{Corresponding author: \texttt{roi.pony@ibm.com}.} \quad Daniel Ezer \quad Adi Raz Goldfarb \\
    \textbf{Idan Friedman} \quad \textbf{Oshri Naparstek} \quad \textbf{Udi Barzelay} \\
    IBM Research Israel}
\fi

\begin{document}
\maketitle

\begin{abstract}
Late-interaction retrieval (ColBERT, ColPali) scores a query against
a document via the \maxsim{} operator. The standard PyTorch
implementation materialises the full query-token $\times$
document-token similarity tensor only to reduce it away. At
ColPali scale this is the single largest tensor in the pipeline
(e.g.\ $21$\,GB in FP16 for $10$K documents) and limits both
candidate set size at inference and batch size during contrastive
training. We present \flash{}~(\flashshort{}), an IO-aware fused GPU kernel that
computes the same \maxsim{} scores without ever materialising the tensor, and
extends the same principle to the training backward. At ColPali scale on A100,
\flashshort{} reduces inference peak memory by $1.4\!-\!2.6\times$ relative
to the deployed chunked baseline ($4.9\!-\!8.9\times$ relative to unchunked
eager execution) and reduces \maxsim{}-operator training memory by two
orders of magnitude, enabling exact reranking over larger resident candidate
pools and contrastive batch sizes that vanilla autograd cannot fit on a
single GPU. The kernel is a drop-in
replacement, \emph{exact up to floating-point evaluation order} under
its stated FP32-accumulation protocol: nDCG@$10$ differs from the
FP32 reference by at most $5{\times}10^{-4}$ on BEIR and REAL-MM-RAG. A
separate INT8 path trades exactness for halved index storage at
high fidelity. Released open-source.\repofn{}
\end{abstract}
\begin{figure*}[t]\centering
\includegraphics[width=0.9\textwidth]{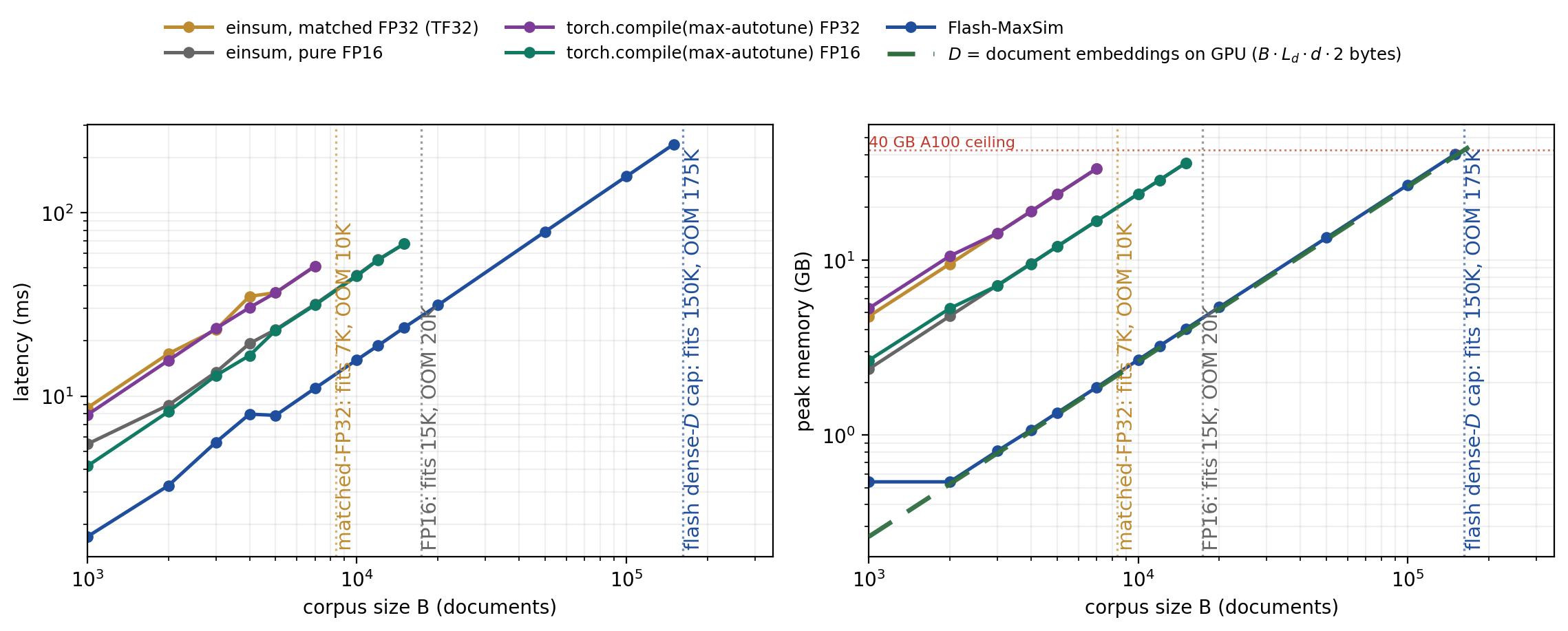}
\caption{ColPali corpus scaling on A100-40GB ($L_q{=}L_d{=}1024$).
Naive einsum (FP16, FP32, both under \texttt{compile-MA}) all OOM
by $B\!\approx\!10\text{--}20$K (markers state last-fitting and
first-OOM $B$; the bench grid steps $7$K$\to$$10$K$\to$$20$K);
\flash{}'s peak tracks the document embeddings (flat
${\sim}0.5$\,GB below $B\!\approx\!2$K, where one-time autotune
scratch dominates) and scales until dense $D$ exceeds VRAM
($B\!\approx\!175$K). Beyond that, OOC streaming
(Tab.~\ref{tab:ooc}) reaches $750$K docs at flat $5.3$\,GB.}
\label{fig:scaling}
\end{figure*}

\section{Introduction}
\label{sec:intro}
Dense single-vector retrieval \cite{dpr,ance,contriever,splade}
scores by one dot product per item; \emph{late-interaction}
retrieval instead keeps a \emph{set} of token-level embeddings and
scores with \maxsim{}:
\begin{equation}
\label{eq:maxsim}
\mathrm{score}(Q, D) = \sum_{i=1}^{L_q} \max_{j=1}^{L_d} \langle Q_i, D_j \rangle,
\end{equation}
with $Q \in \mathbb{R}^{L_q \times d}$, $D \in \mathbb{R}^{L_d \times d}$. This
finer-grained matching is what gives ColBERT \cite{colbert} and its
document-image successor ColPali \cite{colpali} their quality advantage, and
their cost. The natural PyTorch implementation
(Listing~\ref{lst:naive}) forms the $[L_q, L_d]$ similarity
matrix per (query, document) pair via an \texttt{einsum}, then reduces it
($\max$ over document tokens, sum over query tokens). At scale the dominant
memory cost is therefore a $[N_q, B, L_q, L_d]$ tensor ($N_q$
queries against $B$ documents at token lengths $L_q, L_d$), that
exists only to be reduced away.

For visual ColPali ($d\!=\!128$) in the
\emph{page-as-query} regime (full-resolution page images as
\emph{both} query and corpus at $L_q\!=\!L_d\!=\!1024$ patches, as in
visual document-to-document retrieval and de-duplication): a
$10$K-document candidate set makes that tensor $21$\,GB in FP16
($42$\,GB in FP32), OOMing a $40$\,GB GPU before $20$K documents
(Fig.~\ref{fig:scaling}) and capping batch size on $80$\,GB for both
reranking and contrastive training. A short \emph{text} query shrinks
the inference tensor, but the wall returns in in-batch-negatives
training, where $S$ is quadratic in $B$ (\S\ref{sec:bottleneck}). \textbf{The matrix is the
bottleneck, and it never needed to exist.}

\begin{lstlisting}[language=Python,float=t,columns=fixed,
  caption={The standard PyTorch \maxsim{}: the \texttt{[B,Lq,Ld]}
  similarity tensor \emph{S} is materialised in HBM only to be
  reduced away.},label={lst:naive}]
# Q:[Lq,d], D:[B,Ld,d]  (ColPali: Lq=Ld=1024, d=128)
S = torch.einsum("qd,bld->bql", Q, D)  # S:[B,Lq,Ld]; 21 GB at B=10K
scores = S.max(2).values.sum(1)        # [B]; S then discarded
\end{lstlisting}

FlashAttention \cite{flashattention} showed that the analogous attention
matrix can be avoided by IO-aware tiling. We make the same move for
\maxsim{}, which is the headline instance of a \emph{hard-selection}
reduction ($\max$/$\min$/$\arg$ over one axis, sum over another), whose
forward and backward share the same structure across many ML kernels.
We carry the IO-aware principle through the whole structure: the
streaming forward (uniform and ragged corpora), the training backward
that recovers exact gradients, and two serving-oriented extensions (8-bit
quantisation and split-d) built on the same kernels.
Our contributions:
\begin{enumerate}
\item A \textbf{fused forward kernel} (\S\ref{sec:forward}) computing
exact \maxsim{} with peak memory $\approx$ the document embeddings,
never the similarity tensor. It is up to
$\mathbf{3.8\times}$/$\mathbf{4.7\times}$ (A100/H100)
\emph{faster} than naive PyTorch and $0.97\!-\!1.95\times$ vs.\ the
strongest per-cell \texttt{torch.compile} configuration
($2.4\!-\!4.7\times$ vs.\ canonical \texttt{max-autotune})
(Tab.~\ref{tab:fwd}), at up to
$\mathbf{8.9\times}$ \emph{lower peak memory} than unchunked FP16 eager
(Tab.~\ref{tab:fwd_chunked}); this unlocks corpus sizes that OOM naive
on $40$\,GB GPUs. A \texttt{cu\_seqlens} variant adds up to
$\mathbf{4.6\times}$ over padded scoring on ragged corpora, and an FP32-reduction BF16
mode avoids the precision failure of naive bf16 (\S\ref{sec:correct}).
\item A \textbf{fused training backward} (\S\ref{sec:backward}):
inverse-grid CSR (compressed sparse row) for atomic-free destination-owned gradients, plus a
fused \emph{atomic-unified} default at $1.3$--$4.7\times$ less peak
than the CSR variant.
Gradients match FP32 to cosine $1.0000$; batch sizes that OOM
PyTorch are unlocked. A \emph{streamed sharded} variant extends the
same principle to distributed global-negative training with
per-device memory linear in the pool (\S\ref{sec:ddp}): one GPU
outperforms four under conventional cross-device negatives.
\item \textbf{Two serving-oriented extensions}:
\emph{8-bit scoring} (\S\ref{sec:int8}; primarily a $1.97\times$
storage win at Spearman $\rho\!\geq\!0.999$ to the exact scores, with up to a $1.27\times$ speedup
over our own FP16 kernel) and a \emph{split-d forward}
(\S\ref{sec:splitd}) removing the SRAM-spill cliff at $d>512$.
\item \textbf{Evaluation.} Head-to-head against eager PyTorch
(plain and corpus-chunked, the de-facto deployed fix) and
\texttt{torch.compile(max-autotune)}; complete-training-step
memory, equal-data batch-size$\to$quality ($+3.5$ nDCG@$10$ points
at the unlocked $B{=}512$), distributed cross-device-negative, and full
serving-pipeline measurements (\S\ref{sec:ddp},
App.~\ref{app:fullstep}--\ref{app:pipeline}); end-to-end retrieval
parity for the \emph{non-quantised} path verified on BEIR
(ColBERTv2 \cite{colbertv2}) and on four REAL-MM-RAG subsets
(ColPali v1.2 \cite{colpali} + a second public vision-language
encoder, Granite Vision Embedding (GVE)
\cite{granite_vision_embedding}); INT8 separately validated as
high-fidelity but approximate ($\rho\!\geq\!0.999$). Released as
an open-source PyTorch operator for A100/H100 deployments.\textsuperscript{\ref{fn:repo}}
\end{enumerate}

\section{Related Work}
\label{sec:related}
FlashAttention \cite{flashattention,flashattention2} is the
conceptual parent: IO-aware tiling avoiding the materialised
attention matrix, expressed in \texttt{Triton} \cite{triton} and
libraries like \texttt{xFormers} \cite{xformers}.
PagedAttention \cite{vllm} carries the principle into serving.
We apply it to \maxsim{}: a plain running max (idempotent, no
rescaling) and a hard-selection (argmax) backward for training.
Flash-KMeans \cite{flashkmeans} is the closest analog
(materialisation-free online argmin); deltas vs.\ our backward in
App.~\ref{app:related_extended}. ColBERT \cite{colbert} / ColPali
\cite{colpali} define the application; ColBERTv2 \cite{colbertv2}
/ PLAID \cite{plaid} are complementary (residual compression +
centroid-pruned approximate scoring); the ColBERT codebase ships
segmented scoring kernels specialised to that compressed-index
path, while \flash{} targets the dense exhaustive
$L_q\!\times\!L_d\!\times\!d$ rerank those pipelines still invoke
on the surviving candidates. This is a different operating
point, not a head-to-head substitute. Quantisation work targets transformer \emph{weights}
\cite{llmint8,gptq,smoothquant}; \S\ref{sec:int8} instead
quantises the retrieval-side $Q$/$D$ embeddings.
\texttt{torch.compile} \cite{torchinductor} fuses
elementwise ops and is our strongest PyTorch baseline
(App.~\ref{sec:compile}); it cannot remove the materialised
$[B,L_q,L_d]$ einsum output, so all OOM cliffs in
Tab.~\ref{tab:mem} reproduce under \texttt{compile-MA}.

\section{Background and Motivation}
\label{sec:background}

\subsection{\maxsim{} and the Standard GPU Implementation}
A late-interaction model encodes a query as $Q \in \mathbb{R}^{L_q\times d}$ and a
document as $D \in \mathbb{R}^{L_d\times d}$ (token-level, $\ell_2$-normalized,
$d{=}128$; notation summarised in Tab.~\ref{tab:notation}, workload
shapes in Tab.~\ref{tab:shapes}). \maxsim{} (Eq.~\ref{eq:maxsim}) runs at scale in three regimes:
\emph{reranking} (one query vs.\ $B$ docs $\to$ $[B]$),
\emph{multi-query serving} ($N_q$ concurrent user queries against a
shared candidate corpus in a retrieval serving system
$\to$ $[N_q, B]$; App.~\ref{app:multiquery}), and
\emph{in-batch-negatives} training (every query vs.\ every in-batch
document, the same $[N_q, B]$ form). The natural batched implementation materialises
$S = \texttt{einsum}(Q, D) \in \mathbb{R}^{N_q\times B\times L_q
\times L_d}$ in HBM, reads it back for
\texttt{max}$\cdot$\texttt{sum}, and discards it
(App.~\ref{app:background_extended}, Alg.~\ref{alg:naive});
reranking is the $N_q{=}1$ slice.

\subsection{Bottleneck: $S$ Traffic, Roofline, System Limits}
\label{sec:bottleneck}
The footprint of $S$ is $\Theta(B L_q L_d)$. Naive
moves $\Theta(B L_q L_d)$ bytes for $S$ versus $\Theta(B L_d d)$ for
inputs; at ColPali scale (see \S\ref{sec:intro}) the $S$ traffic
exceeds the corpus-side input traffic (the one query amortised) by
$L_q/d\!\approx\!8\times$, and the
materialised $S$ holds $L_q L_d$ entries per (query, document) pair
versus a single output score.
Naive arithmetic intensity sits $4\!-\!5\times$ below the H100
roofline ridge; \flash{} crosses it by removing the $S$ traffic
(Thm.~\ref{thm:io}; App.~\ref{app:background_extended}). The cap is structural: in-batch-negatives training forms an
$[N_q,B,L_q,L_d]$ tensor (ColPali
$B{=}128$ OOMs $80$\,GB; \S\ref{sec:eval}). A second cost compounds
this on variable-length corpora: the dense $[B,L_d,d]$ input must
itself be built by padding every document to the longest $L_d$,
spending memory and compute on tokens that are only masked away,
avoided by the packed \texttt{cu\_seqlens} path (\S\ref{sec:varlen}).

\section{Methodology}
\label{sec:method}

\subsection{Fused Forward: Materialization-Free Scoring via Online Max}
\label{sec:forward}
The forward kernel computes Eq.~\ref{eq:maxsim} without ever writing
$S$ (memory-hierarchy diagram Fig.~\ref{fig:hierarchy} in
App.~\ref{app:background_extended}). It is FlashAttention-style tile-then-reduce, but the in-tile
reduction is a plain \emph{running max}: no log-sum-exp, no
rescaling, because $\max$ is idempotent
(App.~\ref{app:related_extended}).

\begin{algorithm}[t]
\caption{\flash{} forward (one program per (query, document) pair)}
\label{alg:fwd}
\small
\begin{algorithmic}[1]
\State $\textit{acc} \gets 0$ \Comment{FP32 register accumulator}
\For{each query tile $Q_{\mathrm{blk}}$ of $\textsc{b}_q$ rows}
  \State load $Q_{\mathrm{blk}}$ into SRAM \Comment{$[\textsc{b}_q, d]$}
  \State $m \gets -\infty$ \Comment{running max per query token, $[\textsc{b}_q]$}
  \For{each document tile $D_{\mathrm{blk}}$ of $\textsc{b}_d$ rows}
    \State load $D_{\mathrm{blk}}$ into SRAM \Comment{$[\textsc{b}_d, d]$}
    \State $S_t \gets Q_{\mathrm{blk}} D_{\mathrm{blk}}^{\!\top}$ \Comment{$[\textsc{b}_q,\textsc{b}_d]$ in SRAM (FP16 ops, FP32 accumulator)}
    \State mask invalid doc positions to $-\infty$
    \State $m \gets \max(m,\ \mathrm{rowmax}(S_t))$
  \EndFor
  \State $\textit{acc} \gets \textit{acc} + \sum m$
\EndFor
\State write $\textit{acc}$ (one scalar) to HBM
\end{algorithmic}
\end{algorithm}
 Algorithm~\ref{alg:fwd} gives the
per-program pseudo-code.

The kernel launches one program per (query, document) pair, each
emitting a single FP32 scalar; the $[\textsc{b}_q, \textsc{b}_d]$
sub-tile $S_t$ lives only in SRAM and is reduced before the next
document tile overwrites it; the full $[B, L_q, L_d]$ tensor is
never written. Because the sum-of-maxima decomposes over query
chunks, a single kernel handles variable $L_q,L_d$ at runtime, with a
dispatcher selecting specialised variants (ragged, INT8, split-d, and
architecture-specific tile choices; App.~\ref{app:method_extended})
(Alg.~\ref{alg:fwd}).

\begin{proposition}[Exactness]
\label{prop:exact}
Algorithm~\ref{alg:fwd} returns exactly
$\mathrm{score}(Q,D)=\sum_{i=1}^{L_q}\max_{j=1}^{L_d}\langle Q_i,D_j\rangle$,
identical, at matched input dtype and FP32 accumulation, to the dense
reduction up to floating-point evaluation order (agreement with the
TF32-multiply naive baseline of Tab.~\ref{tab:precision} is empirical,
not bitwise),
using only per-program running-max state and $\Theta(N_q B)$ HBM output,
with no HBM-resident $\Theta(B L_q L_d)$ similarity tensor (cf.\ the
naive path, App.~\ref{app:background_extended}, Alg.~\ref{alg:naive}).
Full details in App.~\ref{app:method_extended}.
\end{proposition}

\begin{theorem}[IO complexity]
\label{thm:io}
For $N_q$ queries and $B$ documents at lengths $L_q,L_d$ and
dimension $d$, naive \maxsim{} performs $\Theta(N_q B L_q L_d)$ HBM
accesses and uses $\Theta(N_q B L_q L_d)$
memory. \flash{} (Alg.~\ref{alg:fwd}) performs only
$\Theta(N_q B (L_q{+}L_d) d)$ operand reads and $\Theta(N_q B)$
scalar writes.
The ratio of these asymptotic terms is $L_q/2d$ at $L_q{=}L_d$
($4\times$ at ColPali); the \emph{measured} end-to-end traffic ratio
is larger ($33\times$, Tab.~\ref{tab:traffic}) because the naive path
also writes and re-reads $S$ whereas \flash{} reads each operand once
(proof + arithmetic-intensity analysis in
App.~\ref{app:background_extended}).
\end{theorem}

\noindent\textbf{Ragged corpora and kernel family.} The same
running-max core gates a \texttt{cu\_seqlens} variant for ragged
inputs (\S\ref{sec:varlen}) and the forward family (single-query
/ batched / packed / split-$K$ / INT8) behind a runtime dispatcher
(App.~\ref{app:method_extended}).

\subsection{Inverse-Grid Update: Low-Contention Gradient Aggregation}
\label{sec:backward}
The forward removes a dense value tensor; the backward removes
the dense \emph{gradient} tensor by treating the saved argmax map
as a sparse bipartite routing graph — $\nabla Q$ gathers along
each source's one outgoing edge, $\nabla D$ reduces incoming
edges (Fig.~\ref{fig:bwd_asym}, App.~\ref{app:bwd_algo}).

\subsubsection{Closed-Form Gradients} With upstream gradient
$g_{ij} = \partial L/\partial\,\mathrm{score}(i,j)$ and the forward winner
$t^\star(i,j,s) = \arg\max_t \langle Q_{i,s}, D_{j,t}\rangle$, resolving the $\max$ (a constant
index post-forward) and applying the chain rule gives
\begin{align}
\nabla_{Q_{i,s}} &= \textstyle\sum_j g_{ij}\, D_{j,\,t^\star(i,j,s)}, \label{eq:dq}\\
\nabla_{D_{j,t}} &= \textstyle\sum_{(i,s):\,t^\star(i,j,s)=t} g_{ij}\, Q_{i,s}. \label{eq:dd}
\end{align}
Eq.~\ref{eq:dq} is a collision-free \emph{gather} (one program
per $(i,s)$); Eq.~\ref{eq:dd} is a contended \emph{scatter}
(many sources may pick the same $t$), which is why $\nabla D$ is the hard
side and the inverse-grid CSR exists (Fig.~\ref{fig:bwd_asym},
App.~\ref{app:bwd_algo}).

\paragraph{Backward memory.} Vanilla autograd retains \emph{both}
$S$ and $\nabla S$ at $[N_q,B,L_q,L_d]$; the CSR path saves only
the int32 argmax + a sub-GB transient, and atomic-unified drops the
CSR build too. Idealised totals at ColPali $B{=}128$:
$\mathbf{137}$\,GB (vanilla) / $0.83$\,GB (CSR) /
$\mathbf{67}$\,MB (unified) --- a
$2 L_d = \mathbf{2{,}048\times}$ reduction, independent of $B$
(App.~\ref{app:bwd_mem}).

\subsubsection{Inverse-Grid Kernels} Inside autograd, three GPU
primitives (\texttt{bincount}$\to$\texttt{cumsum},
\texttt{argsort}) invert the saved argmax into a CSR map at
$O(N_q B L_q)$; the $\nabla D$ kernel assigns one program per
destination row, walks its source list, and writes the row
\emph{once}, with no atomics, while $\nabla Q$ is a pure gather. Neither path materialises the $[N_q,B,L_q,L_d]$ gradient. Peak is
essentially embeddings plus the int32 argmax. This yields
Tab.~\ref{tab:train}'s two-orders-of-magnitude
reduction and $B{=}128$ unlock
at $\nabla Q/\nabla D$ cosine $1.0000$ (FP32-cosine-equivalent,
bit equality is not claimed; Alg.~\ref{alg:bwd},
App.~\ref{app:bwd_algo}).

\paragraph{Dispatch.} The dispatcher deterministically selects among
three backward variants
(no autotune trial): \emph{atomic\_unified} (fused
$\nabla Q{+}\nabla D$, the default), plain \emph{atomic}
(low-contention regimes), and \emph{invgrid CSR} (opt-in,
atomic-free destination-owned scatter); criteria and rationale in
App.~\ref{app:bwd_algo}. The same machinery ports to Chamfer
distance at $77\!-\!152\times$ (App.~\ref{sec:chamfer});
evidence of transfer, not a coverage claim.

\subsection{Quantised Scoring}
\label{sec:variants}
\label{sec:int8}
A serving-oriented INT8 variant per-token symmetric-quantises $Q$ and $D$
(absmax-based, computed at index time for $D$ and on-the-fly for
$Q$; no separate calibration pass), fuses dequant into the kernel,
and dispatches INT8 tensor cores. It halves $D$-storage
($\approx\!1.97\times$ with the FP16 per-token scale) and is
$\mathbf{3\!-\!6\times}$ over the dequant-then-naive path
(over FP16 \flash{} itself: a modest $1.07\!-\!1.27\times$ across
the ColPali $B{=}128$--$10$K sweep, $0.93\!-\!1.27\times$ across the
five canonical shapes, Tab.~\ref{tab:int8}).
The win is a \emph{deferred-dequant} pattern: the per-row scale
is applied once per output cell, not per input element
(Alg.~\ref{alg:int8}, App.~\ref{app:int8_algo}). INT8 is the only
approximate \flash{} variant ($\rho\!\geq\!0.999$, top-$20$
$\geq\!95\%$; Tab.~\ref{tab:int8},\ref{tab:int8_B}).

\begin{table}[t]\centering\small
\caption{Forward speedup at $B{=}1$K, $N_q{=}1$, measured under
the full protocol of App.~\ref{app:bench_protocol} in one campaign
per GPU (interleaved, L2-flushed, CUDA-event medians; IQR
$\leq\!1.2\%$ of the median at every cell). First two columns:
\flash{} vs.\ naive at matched precision (FP32 accumulation,
TF32) on A100 / H100. Third: vs.\ the \emph{strongest}
\texttt{torch.compile} configuration per cell
(\texttt{max-autotune-no-cudagraphs} on the FP16 einsum, A100;
Tab.~\ref{tab:nocg}); the weaker canonical graphs-on
\texttt{max-autotune} comparison ($2.4\!-\!4.7\times$) is audited in
App.~\ref{sec:compile}. Absolute values: Tab.~\ref{tab:fwd_abs}. Source:
\texttt{bench\_protocol\_grade\_shapes\_*.json}.}
\label{tab:fwd}
\setlength{\tabcolsep}{4pt}
\begin{tabular}{lccc}
\toprule
Shape ($L_q,L_d$) & A100$\times$ & H100$\times$ & compile-best \\
\midrule
textual ($32,300$)     & $1.2\times$ & $1.1\times$ & $0.97\times$ \\
long-doc ($32,1024$)   & $1.9\times$ & $1.7\times$ & $1.21\times$ \\
medium ($128,1024$)    & $2.8\times$ & $3.1\times$ & $1.73\times$ \\
visual ($512,1024$)    & $3.4\times$ & $4.0\times$ & $1.83\times$ \\
ColPali ($1024^2$)     & $\mathbf{3.8\times}$ & $\mathbf{4.7\times}$ & $\mathbf{1.95\times}$ \\
\bottomrule
\end{tabular}
\end{table}

\section{Experiments}
\label{sec:eval}

\paragraph{Target deployment.}
The evaluation targets the operating points where \maxsim{} is the
throughput- or memory-limiting operator: $1$K--$100$K-candidate
reranking, in-batch-negatives training, ragged-corpus scoring, and
multi-query serving (App.~\ref{app:multiquery}: up to $45$K
queries/s on textual, and $4.5\times$ over textbook-FP32 naive at
the ColPali $B{=}128$ serving microbenchmark).

\subsection{Setup}
We evaluate on A100-SXM4-80GB (H100 for the forward sweep; A100-40GB
for the corpus-scaling stress test, Fig.~\ref{fig:scaling}) with
PyTorch 2.8 + Triton 3.5/3.6, TF32 on. Numbers are CUDA-event
medians of $50$ post-warmup runs, with autotune/JIT and casts
outside the timed region; run-to-run IQR is $\leq\!1.2\%$, so we omit
interval bars. Tables ~\ref{tab:fwd}, \ref{tab:fwd_chunked} are \emph{inference mode}, the training
forward is timed in Tab.~\ref{tab:train}. Five canonical shapes
(\textit{textual}, \textit{long-doc}, \textit{medium},
\textit{visual}, \textit{ColPali}; $L_q,L_d$ from $32,300$ to
$1024,1024$ at $d{=}128$, $N_q{=}1$) span the deployed operating
points. The full shape/encoder map, precision protocol, baseline
glossary, and fairness statement are in App.~\ref{app:setup}
(Tab.~\ref{tab:shapes},~\ref{tab:precision},~\ref{tab:baselines}).

\subsection{Baselines}
Latency uses three baselines, each in matched-FP32 and pure-FP16
variants where applicable: \emph{naive eager}
(\texttt{einsum}+\texttt{max}+\texttt{sum}); \emph{chunked FP16
eager} (the production-default no-compile fix, oracle chunk-size;
App.~\ref{app:chunked_eager}); and
\texttt{torch.compile(max-autotune)}. INT8 adds a
\emph{dequant-then-naive} baseline, and correctness uses a true-FP32
reference (full: App.~\ref{app:setup},
Tab.~\ref{tab:baselines}).

\begin{table}[t!]\centering\small
\caption{Inference frontier (shapes defined in
Tab.~\ref{tab:shapes}), A100-80GB, full protocol, one
campaign. pk$_{\text{F/c/van}}$ = peak GPU memory for \flash{}\,/\,%
chunked FP16 eager (chunk $1024$, the
sweep's best; App.~\ref{app:chunked_eager})\,/\,unchunked FP16
eager. All baselines fit at every cell; the
win is $2.6\times$ latency at $1.4\!-\!2.6\times$ (vs.\ chunked) and
$4.9\!-\!8.9\times$ (vs.\ unchunked) lower peak, not feasibility.
Speedups here are vs.\ the \emph{pure-FP16} chunked baseline; the
larger Tab.~\ref{tab:fwd} speedups are vs.\ FP32-accumulation naive
at $B{=}1$K.}
\label{tab:fwd_chunked}
\setlength{\tabcolsep}{2.6pt}
\begin{tabular}{cccccc}
\toprule
shape & $B$ & speedup & pk$_{\text{F}}$ & pk$_{\text{c}}$ & pk$_{\text{van}}$ \\
\midrule
visual   & 10K & $2.61\times$ & $2.7$\,GB & $4.9$\,GB & $13.4$\,GB \\
visual   & 20K & $2.64\times$ & $5.4$\,GB & $7.5$\,GB & $26.6$\,GB \\
ColPali  & 10K & $2.64\times$ & $2.7$\,GB & $7.0$\,GB & $24.0$\,GB \\
ColPali  & 20K & $2.66\times$ & $5.4$\,GB & $9.7$\,GB & $47.8$\,GB \\
\bottomrule
\end{tabular}
\end{table}

\subsection{Forward Latency and Memory}

Two scenarios anchor the section. \emph{Inference rerank}: $10$K
ColPali candidates in $\mathbf{16.2}$\,ms at $\mathbf{2.7}$\,GB
vs.\ chunked FP16 eager's $42.8$\,ms at $7.0$\,GB. That is
$\mathbf{2.64\times}$ the speed at $2.59\times$ lower peak
($8.9\times$ vs.\ unchunked) and the ratio is flat across the sweep
(Tab.~\ref{tab:fwd_chunked}). On $40$\,GB the gap becomes
feasibility (Fig.~\ref{fig:scaling}).
\emph{Contrastive training}: \flash{} runs $B{=}128$ in-batch
negatives at $0.39$\,GB operator peak on a single A100-80GB, where
naive autograd OOMs and operator-level checkpointing pays
$8\times$ the step time at ${\sim}45\times$ the peak
(Tab.~\ref{tab:train}). The rest of the section breaks each down
by shape, baseline, and corpus size.

As shown in Table~\ref{tab:fwd}, \flash{} is faster at every shape
over naive, up to $3.8\times$ (A100) / $4.7\times$ (H100) at ColPali,
and $0.97\!-\!1.95\times$ vs.\ the \emph{strongest} per-cell
\texttt{torch.compile} configuration
(\texttt{max-autotune-no-cudagraphs}, FP16): parity only at the
launch-bound textual shape, ${\approx}2\times$ at the memory-bound
cells. Canonical \texttt{compile-MA} trails further, by
$2.4\!-\!4.7\times$; both flavours and their OOM behaviour are
audited in App.~\ref{sec:compile} (Tab.~\ref{tab:nocg}).

\paragraph{HBM traffic + robustness.} Any exact scorer must read
the $Q,D$ operands and write the scores. \flash{} approaches this
operand-read floor on the corpus side ($0.26$\,GB of $D$ traffic vs.\
naive's $8.65$\,GB of $S$ traffic at ColPali $B{=}1$K, with $Q$ loaded
once; full accounting in Tab.~\ref{tab:traffic}) and the matmul
compute floor ($1.70$ vs.\ $1.72$\,ms). Tile sweeps are flat within $3\%$
(robustness in App.~\ref{sec:compile}).

\paragraph{Two-axis benefit.} The benefit has two \emph{independent}
axes. \textbf{Latency}: where both fit, \flash{} processes each
candidate in $\mathbf{1.6}$\,µs vs.\ naive's $4.5$\,µs and
chunked-eager's $4.3$\,µs, a $\mathbf{2.6\!-\!2.85\times}$ steady-state
reduction from removing the $S$ traffic (Thm.~\ref{thm:io}).
\textbf{Memory headroom}: \flash{}'s peak grows with the embeddings
($\Theta(B L_d d)$), not $S$ ($\Theta(B L_q L_d)$); on $80$\,GB this
is $1.4\!-\!8.9\times$ headroom over chunked\,/\,unchunked eager
(Tab.~\ref{tab:fwd_chunked}), and on $40$\,GB it becomes feasibility
(naive OOMs by $B{=}150$K where \flash{} still runs,
Tab.~\ref{tab:twoaxis}). Headroom is
never a throughput multiplier: past saturation total time is linear in
$B$, so the per-candidate ratio is unchanged.

\noindent For \emph{training}, axis 2 is decisive: in-batch
negatives at $B{=}128$ score all pairs in one step (plain gradient
accumulation does not reproduce the same in-step in-batch negative set
without extra machinery such as cross-batch memory or cached
embeddings); naive autograd OOMs, and the
chunked-recompute alternative pays $8\times$ the step time
(measured below); quantified in Tab.~\ref{tab:train}.

\paragraph{Ragged corpora.} Forming the dense $[B,L_d,d]$ document
tensor that \texttt{einsum} consumes already forces every document to
be padded to the longest $L_d$ in the batch, spending memory and
compute on padding tokens that are only masked away afterwards, a
cost the dense baseline pays before $S$ even materialises. The
\texttt{cu\_seqlens} variant (\S\ref{sec:varlen}) instead packs only
real tokens: up to $\mathbf{4.6\times}$ on highly-ragged corpora,
$4.3\times$ at a HotpotQA-like distribution, exact scores
(Tab.~\ref{tab:varlen}).

\paragraph{Out-of-core corpus scaling.}\label{sec:ooc}
Corpora beyond VRAM stream from host RAM: one ColPali query
vs.\ $750$K docs ($197$\,GB, $2.3\times$ VRAM) runs at flat
$5.3$\,GB and $52$K\,docs/s (Tab.~\ref{tab:ooc}).

\subsection{Training Step} At ColPali in-batch negatives
(Tab.~\ref{tab:train}) the naive backward materialises the
$[B,B,L_q,L_d]$ tensor \emph{and} its gradient;
\flash{}'s backward (\S\ref{sec:backward}) removes both. The dominant win is the batch-size unlock: $B{=}128$ OOMs naive
on $80$\,GB while \flash{} (unified default) trains at
$0.39$\,GB peak; at $B{=}64$ where both fit, \flash{} is
$6.1\times$ faster with over $\mathbf{200\times}$ less
peak memory. Gradients match FP32 to $\nabla Q / \nabla D$ cosine
$1.0000$; a $500$-step end-to-end contrastive loop at $B{=}64$
reproduces the loss trajectory of an FP16-in-dtype naive baseline
(matched FP32 OOMs once model and optimizer state are resident,
unlike the bare operator step of Tab.~\ref{tab:train}) within
$1.4{\times}10^{-3}$ max per-step drift
(full parity curve in
Fig.~\ref{fig:training_parity}, App.~\ref{app:training_parity}).

\begin{table}[t]\centering\small
\caption{Contrastive training step (fwd+bwd) at the \maxsim{}-operator
level, ColPali $L_q{=}L_d{=}1024$, A100-80GB, backward path labeled
per row (\emph{unified} = atomic-unified, the dispatcher default;
\emph{CSR} = opt-in atomic-free inverse-grid; \emph{recompute}
= operator-level checkpointing, block size swept, best shown).
Naive:
\texttt{einsum}+\texttt{max}+\texttt{sum} through autograd at
matched precision. Synthetic $Q,D$; encoder activations, optimizer
state, weight gradients \emph{not} included; peak is the per-step
transient the \maxsim{} backward dominates (byte budget:
App.~\ref{app:bwd_mem}; complete-step measurements incl.\ encoder and
optimizer: App.~\ref{app:fullstep}). Source:
\texttt{bench\_protocol\_grade\_train\_*.json}}
\label{tab:train}
\setlength{\tabcolsep}{4pt}
\begin{tabular}{llccc}
\toprule
$B$ & path & step ms & peak & vs naive \\
\midrule
$64$  & naive            & $81.3$ & $51.8$\,GB & ---  \\
      & recompute        & $109.2$ & $8.8$\,GB & $0.74\times$ \\
      & \flashshort{} unified & $\mathbf{13.4}$ & $\mathbf{0.24}$\,GB & $6.1\times$, $\mathbf{217\times}$ mem \\
      & \flashshort{} CSR     & $11.7$ & $0.39$\,GB & $6.9\times$ \\
$128$ & naive            & \textbf{OOM} & \textbf{OOM} & --- \\
      & recompute        & $426.6$ & $17.5$\,GB & feasible \\
      & \flashshort{} unified & $\mathbf{53.6}$ & $\mathbf{0.39}$\,GB & $8.0\times$ vs.\ rec. \\
      & \flashshort{} CSR     & $46.0$ & $1.03$\,GB & $9.3\times$ vs.\ rec. \\
\bottomrule
\end{tabular}
\end{table}

\paragraph{Backward-path choice.}
CSR is $13\!-\!17\%$ faster at these A100 cells
(Tab.~\ref{tab:train}) but atomic-unified holds $1.3\!-\!4.7\times$
less peak ($2.6\times$ at A100 $B{=}128$) and stays flat in $B$, so it
is the dispatcher default for in-batch negatives; H100 ablation,
contention caveat, and CSR-build decomposition in
App.~\ref{app:bwd_ablations}, Tab.~\ref{tab:bwd_abl}. The
chunked alternative to \flash{}'s backward is analysed in
App.~\ref{app:bwd_ablations}.

\paragraph{Why operator-level.} We isolate \maxsim{} because its
transient memory scales as $B^2 L_q L_d$: encoder, optimizer, and
weight-gradient memory are shared across \emph{any} scorer, whereas
the materialised similarity tensor is unique to the naive path and is
exactly what \flash{} removes. Operator-level is also the per-step cost
of the scorer substep in cached-encoding
(GradCache-style~\cite{gradcache}) large-batch
recipes, where embeddings are encoded in micro-batches and the scorer
dominates large-batch memory. Complete-step measurements including
encoder activations and optimizer state are in App.~\ref{app:fullstep};
at equal data, the batch sizes this unlocks improve retrieval quality
(App.~\ref{app:batchquality}), and the sharded variant changes the
per-device memory scaling of distributed global-negative training
(\S\ref{sec:ddp}). Tab.~\ref{tab:e2e_train} summarises the three
end-to-end results: the operator saving is what moves the
complete-step wall at large $B$, the unlocked batch measurably
improves the retriever at equal data, and one GPU under the streamed
sharded loss outperforms four under conventional cross-device
negatives.

\begin{table}[h]\centering\small
\caption{End-to-end training summary (A100-80GB, ColBERT/BERT-base;
complete steps incl.\ encoder and optimizer; $83.1$\,GB
$=77.4$\,GiB, fits the $80$-GiB card). ``Ref'' is the
reference scorer in identical trainer code, or the differentiable
all-gather loss (last row). Quality row: equal data, $5$ seeds per
point. Full tables and protocols:
App.~\ref{app:fullstep}--\ref{app:ddp}.}
\label{tab:e2e_train}
\setlength{\tabcolsep}{2.5pt}
\resizebox{\columnwidth}{!}{%
\begin{tabular}{l l l}
\toprule
measurement & ref & \flashshort{} \\
\midrule
step peak, $B{=}512$ & OOM & $\mathbf{83.1}$\,GB \\
step peak, ckpt, $B{=}1024$ & $67.9$\,GB & $\mathbf{26.4}$\,GB \\
nDCG@$10$, $B\,64{\to}512$ & OOM at $512$ & $\mathbf{+3.5}$\,pts \\
global negs., pool $2048$ & $985$\,ms, $4$ GPU & $\mathbf{230}$\,ms, $1$ GPU \\
\bottomrule
\end{tabular}}
\end{table}

\subsection{Distributed Training with Global Negatives}
\label{sec:ddp}
Cross-device negatives are the standard multi-GPU route to large
negative pools, but the conventional differentiable all-gather
still scores each device's local queries against the full global
pool: with $W$ GPUs and a global pool of $B$ documents, per-device
memory holds an $O(B^2/W)$ score block, so a fixed memory budget
grows the reachable pool only as $\sqrt{W}$.
Our \emph{streamed sharded loss} computes the same objective --- the
standard InfoNCE cross-entropy, whose normaliser is a softmax over
all $B$ pool scores --- but shards the documents, communicates
the short query side, and recomputes \maxsim{} tiles under a
two-level online softmax (per-tile, then across ranks), keeping
per-device memory \emph{linear} in $B$.

Measured at $W{=}1,2,4$ (A100-80GB, ColBERT shapes), conventional
follows the quadratic law: $72.8$\,GB per rank at pool $B{=}2{,}048$
on four GPUs, OOM at $4{,}096$. The streamed loss is
$4.2\!-\!17\times$ faster at matched $W{=}4$ and reaches a
$65{,}536$-negative pool at $14.0$\,GB/device. Most directly: at
$B{=}2{,}048$, \emph{one} GPU under the streamed loss is
$\mathbf{4.3\times}$ faster than conventional on \emph{four}
($230$ vs.\ $985$\,ms) at $85\times$ less per-device memory
($0.86$ vs.\ $72.8$\,GB). Gradients match the reference at every
cluster size (cosine $\geq 0.9999998$ on $\nabla Q,\nabla D$).
Since quality improves monotonically with pool size at equal data
(App.~\ref{app:batchquality}), this turns negative-pool scale from
a GPU-count question into a kernel question. Full grid and
protocol: App.~\ref{app:ddp}.

\subsection{Numerical Correctness}\label{sec:correct}
At the operator level FP32 accumulation gives max relative error
$4{\times}10^{-7}$ at ColPali ($\rho{=}0.999999$, $100\%$
top-$20$/$50$ overlap), confirming Prop.~\ref{prop:exact};
\flashshort{}'s BF16-input mode, which accumulates in FP32 inside
the kernel, is $35\!-\!87\times$ tighter and keeps
rankings stable where naive bf16 flips them
(App.~\ref{app:method_extended}). End-to-end, \flash{} matches FP32
nDCG@$10$ \emph{to four decimals} with $100\%$ top-$K$ overlap on
BEIR \cite{beir} (ColBERTv2, text) and REAL-MM-RAG~\cite{realmmrag}
(ColPali, vision); with a second encoder (GVE), nDCG@$10$ is within
$5{\times}10^{-4}$, Recall@$10$ within $1.2{\times}10^{-3}$, and
top-$10$ overlap $\geq\!99.89\%$ (Tab.~\ref{tab:beir_ndcg},
\ref{tab:colpali_ndcg}). Parity is verified at corpus scale on
HotpotQA-$500$K; the same exact out-of-core path scales to NQ at
$2.68$M documents, where the dense reference cannot run
(App.~\ref{app:beir_scale}, \ref{app:colpali_ndcg_full}). The INT8 path, the only approximate one,
stays within $\mathbf{|\Delta|\!\leq\!0.003}$ nDCG@$10$ end-to-end
($\rho\!\geq\!0.999$, top-$20$ $\geq\!95\%$; Tab.~\ref{tab:int8},
\ref{tab:int8_realmmrag}).

\FloatBarrier
\section{Conclusion}
\flash{} removes the $[N_q,B,L_q,L_d]$ similarity tensor that
bottlenecks \maxsim{} by carrying FlashAttention's IO-aware tiling
through the \emph{whole} operator: a materialisation-free forward and
an exact training backward that never forms $\nabla S$. The result is
a drop-in PyTorch operator that cuts inference peak memory by
$1.4\!-\!2.6\times$ vs.\ deployed chunking ($4.9\!-\!8.9\times$ vs.\ unchunked
eager) and operator-level training peak by two orders of magnitude (batch
sizes that OOM one GPU now fit), is faster at every shape over naive
(up to $3.8\times$/$4.7\times$; $0.97\!-\!1.95\times$ vs.\ the
strongest per-cell \texttt{torch.compile} configuration,
$2.4\!-\!4.7\times$ vs.\ canonical \texttt{max-autotune}), scores
ragged corpora without padding (up to $4.6\times$ over padded
scoring), and preserves nDCG@$10$ to four decimal
places on BEIR and ColPali REAL-MM-RAG; an INT8 path
halves index storage. The win is the hard-selection structure
($\max$/$\min$/$\arg$ over one axis, sum over another): a reusable
template beyond \maxsim{}. Released
open-source.\textsuperscript{\ref{fn:repo}}

\newpage
\section*{Limitations}
\flash{} helps when the similarity traffic $L_q L_d$ dominates
the operand traffic $(L_q{+}L_d)\,d$; at small launch-bound shapes
($L_q,L_d\!\lesssim\!64$) it is at parity, and a compiled FP16 einsum
can be up to ${\sim}1.4\times$ faster at tiny independent-pair batches
($N\!\leq\!16$). The inverse-grid CSR backward offers little benefit
when atomic contention is already
near zero (expected sources per destination near one, i.e.\ a small
source set relative to the destination space). \flash{} keeps only
the single best match per query token; scoring variants that
aggregate the $k\!>\!1$ best matches would need a different kernel.
Single-step training speedup is modest at matched
precision; the dominant training win is memory and the batch-size
unlock. Hardware coverage is A100/H100 only; the portable Triton
kernels may need tile re-tuning on other architectures (consumer
NVIDIA, Volta/Turing, AMD MI), where the dispatcher falls back to
eager PyTorch. Exactness is relative to the precision protocol
(Tab.~\ref{tab:precision}): the same score up to floating-point
reassociation under matched dtype and FP32 accumulation, not bitwise
equality to a specific PyTorch kernel. With $S$ removed,
large-candidate reranking is bound instead by embedding storage,
host-device streaming (\S\ref{sec:ooc}), candidate generation, and
encoder latency. Our serving measurements (App.~\ref{app:pipeline})
use an exact single-vector first stage to hold candidate selection
fixed; an end-to-end study inside compressed-index systems
(PLAID/ANN) is future work. Complete-training-step gains
(App.~\ref{app:fullstep}) depend on where the scorer crossover falls:
encoder-bound configurations (e.g.\ single-GPU ColPali text-query
fine-tuning at small $B$) see no memory benefit from any scorer
choice.

\bibliography{references}

\onecolumn
\appendix
\section{Benchmarking Protocol: Flash vs.\ \texttt{compile} vs.\ Eager}
\label{app:bench_protocol}

Protocol used in full for the headline tables
(Tabs.~\ref{tab:fwd}, \ref{tab:fwd_abs}, \ref{tab:fwd_chunked},
\ref{tab:nocg}, \ref{tab:train}); secondary appendix sweeps apply
the precision, warmup, and per-method-memory rules with
warm-cache sequential timing:

\begin{itemize}\setlength\itemsep{1pt}
\item \textbf{Precision tiers.} Report eager separately per tier:
pure FP16; matched precision (FP32 accumulation, TF32 on); true
FP32 (TF32 off) as correctness reference only. All
FP16$\to$FP32 casts hoisted outside the timed region.
\item \textbf{Compile flavours.} \texttt{max-autotune} graphs-on
and \texttt{-no-cudagraphs}, each over FP16 and FP32; report the
strongest per cell. $\geq\!100$ warmup calls so autotune and graph
capture settle; autotune in a subprocess
(\texttt{torch.\_inductor.config.autotune\_in\_subproc}).
\item \textbf{Timing.} CUDA-event (or synced
\texttt{perf\_counter}) medians; methods interleaved round-robin
within one loop to cancel clock/thermal drift; L2 cache flushed
($100$\,MB write) before each timed call for cold-cache numbers.
\item \textbf{Dispersion.} Report IQR; within-campaign IQR
($\leq\!1.2\%$ here) does not bound cross-campaign drift
(${\sim}17\%$ on millisecond cells); compute every ratio within
one campaign.
\item \textbf{Memory.} \texttt{reset\_peak\_memory\_stats} per run;
one method per process, because CUDA-graph private pools and prior
methods' operand copies survive \texttt{empty\_cache()}. For OOM
claims, record the exception text (``tried to allocate
$X$\,GiB'').
\item \textbf{Baselines.} Always include the deployed workaround,
chunked FP16 eager with the chunk size swept
(App.~\ref{app:chunked_eager}), alongside eager and compile.
\item \textbf{Validation.} Operator-level error plus end-to-end
retrieval metrics on real qrels; cross-check ratios with an
independently implemented harness (App.~\ref{sec:compile}).
\end{itemize}

\section{Comparison to \texttt{torch.compile}}
\label{sec:compile}
Our main results report speedups against eager naive PyTorch at matched precision
(\S\ref{sec:eval}). For completeness we summarise an independent comparison
against \texttt{torch.compile(mode="max-autotune")} of the same naive
einsum-then-max baseline, on A100-80GB at the shapes of Table~\ref{tab:fwd} and
Table~\ref{tab:mem}.

\paragraph{Protocol.} PyTorch 2.8 + Inductor (CUDA 12.x driver
555.x). \texttt{compile-MA} is invoked as
\texttt{torch.compile(f, mode="max-autotune", dynamic=False)},
which by default enables CUDA-graph capture via
\texttt{cudagraph\_trees}; the private-pool inflation discussed
below comes from this default. Disabling graphs
(\texttt{mode="max-autotune-no-cudagraphs"}) removes the inflation
but \emph{also} removes the autotuner's strongest mode, so
neither flavour is uniformly fastest; Tab.~\ref{tab:fwd}'s third
column reports the strongest per-cell configuration (graphs-off on
the FP16 einsum, summarised in Tab.~\ref{tab:nocg} below), while the
canonical graphs-on matched-FP32 comparison is reported in this
appendix.
\texttt{torch.set\_float32\_matmul\_precision('high')} (TF32 on) is
set for both compile-MA and the eager baselines. The per-cell
\emph{audit} latencies quoted in the ``Forward dense'' paragraph
below (the \texttt{bench\_compile\_ma\_audit} JSON; warm-up $5$ calls,
median of $10$ CUDA-event measurements, input cast hoisted) are a
\emph{legacy warm-cache campaign} and are \emph{not} used for any
headline ratio. The strongest-configuration table
(Tab.~\ref{tab:nocg}) and all headline tables use the full protocol
of App.~\ref{app:bench_protocol}.

\paragraph{Forward dense.} At ColPali $B{=}1$K on A100-80GB,
\texttt{compile-MA} (canonical graphs-on) runs at $6.8$\,ms vs.\
\flash{}'s $1.77$\,ms (a $\mathbf{3.9\times}$ gap) and is
slightly slower than eager FP32 ($6.6$\,ms) at this shape: the
CUDA-graph private pool inflates per-call overhead at small $B$. At ColPali $B{=}10$K, \texttt{compile-MA} OOMs on $80$\,GB:
its private graph pool must pre-allocate the $40$\,GB FP32
$[B,L_q,L_d]$ intermediate, which together with eager-path residuals
exceeds the budget. Eager FP32 (no graph pool) runs the same shape
in $67$\,ms with $50$\,GB peak; \flash{} runs it in $16$\,ms with
$2.7$\,GB peak. (Latencies from
\texttt{data/bench\_compile\_ma\_audit\_NVIDIA\_A100-SXM4-80GB.json};
\flash{} peak from the fresh-process measurement of
Tab.~\ref{tab:fwd_chunked} — the audit JSON's in-process
\texttt{peak\_gb} for \flash{} at this cell ($7.9$\,GB) includes the
preceding FP32 baseline's still-resident operand copies and is an
allocator artifact, not \flash{}'s footprint.)

\paragraph{Memory and OOM.} \texttt{compile} does not remove the
materialised $[B,L_q,L_d]$ intermediate; its private graph pool
only adds overhead. Naive FP32 OOMs by $B{=}20$K on $80$\,GB and
\texttt{compile-MA} OOMs earlier (already at $B{=}10$K) due to the
graph-pool reservation — a failure mode specific to the canonical
graphs-on FP32 flavour: the graphs-off FP16 flavour has no pool
and fits through $B{=}20$K at $47.3$\,GB (Tab.~\ref{tab:nocg}),
mirroring the chunked-eager picture of
App.~\ref{app:chunked_eager}. \flash{} uses $\approx\!8\times$
less memory wherever both fit and continues to scale linearly past
(Table~\ref{tab:mem}, Fig.~\ref{fig:scaling}).

\paragraph{Strongest compile configuration per cell.}
Tab.~\ref{tab:nocg} reports
\texttt{max-autotune-no-cudagraphs} applied to the \emph{FP16}
einsum — empirically the fastest PyTorch configuration we found at
these shapes (it beats canonical graphs-on compile-MA at every
cell measured, e.g.\ $3.44$ vs.\ $6.82$\,ms at ColPali $B{=}1$K,
Tab.~\ref{tab:fwd_abs}).
Against this strongest configuration \flash{}'s margin is
$0.97\!-\!1.95\times$: parity at the launch-bound textual shape
(consistent with the Limitations note) and ${\approx}2\times$
at the memory-bound cells — converging to the FP16-eager tier of
App.~\ref{app:chunked_eager}, as expected, since no compile mode
removes the $S$ traffic that Thm.~\ref{thm:io} eliminates.

\begin{table}[h]\centering\small
\caption{\texttt{compile(max-autotune-no-cudagraphs)} of the FP16
einsum — the strongest PyTorch configuration per cell — vs.\
\flash{}, A100-80GB, measured in the full-protocol campaign used for
Tab.~\ref{tab:fwd} (not the legacy audit paragraph above). In a
separate feasibility run this flavour has no graph pool and fits
through ColPali $B{=}20$K ($47.3$\,GB peak,
\texttt{bench\_compile\_ma\_nocg\_*.json}).}
\label{tab:nocg}
\setlength{\tabcolsep}{4pt}
\begin{tabular}{lrrr}
\toprule
shape & nocg ms & \flash{} ms & $\times$ \\
\midrule
textual   & $0.20$ & $0.20$ & $0.97$ \\
long-doc  & $0.37$ & $0.31$ & $1.21$ \\
medium    & $0.66$ & $0.38$ & $1.73$ \\
visual    & $1.79$ & $0.98$ & $1.83$ \\
ColPali   & $3.44$ & $1.77$ & $1.95$ \\
\bottomrule
\end{tabular}
\end{table}

\paragraph{Manual chunking with compile.} The strongest user-side workaround,
manually chunking the corpus and \texttt{compile-MA}-ing each chunk, lands at
$37.4$\,ms with $3.05$\,GB at $B{=}10$K, still $2.3\times$ slower than \flash{}.

\paragraph{Variable-length.} \texttt{compile} cannot rewrite the user's padding
strategy. Against \texttt{compile-MA} of the padded baseline, \flash{}'s
\texttt{cu\_seqlens} variant is up to $3\times$ faster on ragged corpora.

\paragraph{KernelBench-style~\cite{kernelbench} \texttt{fast\_p} summary.} Across all $53$ cells in
the audit ($5$ dense forward shapes, $6$ corpus-scaling values of $B$, $6$
long-doc values of $B$, $4$ varlen distributions, $4$ chunked-compile shapes,
plus ablations), \flash{} achieves \texttt{fast\_1} (positive speedup)
over the best of eager, chunked, and \emph{canonical} \texttt{compile-MA}
on $\mathbf{53/53}$ cells, \texttt{fast\_2} on $38/53 = 72\%$ of cells, and
\texttt{fast\_3} on $18/53 = 34\%$; a further $8/53 = 15\%$ are
OOM-unlocks where \texttt{compile-MA} cannot run at all. Against the
\emph{per-cell strongest} configuration
(\texttt{compile-MA-no-cudagraphs}, Tab.~\ref{tab:nocg}) \flash{} is
strictly faster on every memory-bound cell and at parity ($0.97\times$)
on the single launch-bound textual cell, consistent with the
Limitations.

\noindent In summary, \texttt{torch.compile} narrows the dense-forward latency
gap to roughly $2\times$, but is compile-invariant on the three properties that
motivate this work: peak memory, the OOM frontier, and padding-free scoring.

\section{Chunked FP16 Eager: the Production-Default Baseline}
\label{app:chunked_eager}

\begin{table}[h]\centering\small
\caption{Two-axis benefit, ColPali on A100-40GB. Cells are
single-call latency. Axis~1 (per-candidate latency) is constant at
$\sim\!2.85\times$ within the fits row; axis~2 (memory) determines
which row is reachable at all. Flash dense caps at $B\!\approx\!175$K
(out-of-core streaming lifts this, \S\ref{sec:ooc}).}
\label{tab:twoaxis}
\begin{tabular}{lcc}
\toprule
                  & naive FP16    & \flash{}              \\
\midrule
$B{=}15$K  (fits) & $68$\,ms      & $\mathbf{24}$\,ms     \\
$B{=}150$K (naive OOMs)
                  & \textbf{OOM}  & $\mathbf{240}$\,ms    \\
\bottomrule
\end{tabular}
\end{table}

The most common deployed fix for the similarity-tensor memory wall is
neither \texttt{torch.compile} nor a fused kernel: it is chunking the
corpus and running the plain FP16 eager einsum per chunk. We sweep
chunk sizes $\{1, 4, 16, 64, 256, 1024, 4096, 16384\}$ at every cell
and report the \emph{best} (lowest-median-latency, non-OOM)
configuration, i.e.\ the baseline is granted oracle chunk-size tuning
that a deployment would have to find empirically.
Tab.~\ref{tab:chunked_sweep} reports all cells; the frontier subset
appears as Tab.~\ref{tab:fwd_chunked} in the main text.

Two observations. First, \flash{}'s win over this baseline is a flat
$\mathbf{2.6\!-\!2.7\times}$ at every frontier cell — consistent with
the per-candidate analysis around Tab.~\ref{tab:twoaxis}, since
chunking changes neither total $S$-traffic nor precision. Second, on A100-80GB even the \emph{unchunked} FP16 eager einsum
fits through ColPali $B{=}20$K ($47.8$\,GB peak). On large-memory
GPUs the inference-side benefit of \flash{} is therefore the
latency and the $1.4\!-\!8.9\times$ peak-memory headroom, not
feasibility; the
feasibility framing applies to $40$\,GB-class GPUs
(Fig.~\ref{fig:scaling}) and to training (Tab.~\ref{tab:train}),
where no chunking strategy avoids re-materialising the gradient of
the similarity tensor.

\paragraph{Cost per scored candidate.} The latency ratio converts
directly to GPU-hours: at ColPali shape, \flash{} spends
$1.6$\,µs per candidate vs.\ chunked FP16 eager's $4.3$\,µs, i.e.\
$0.44$ vs.\ $1.19$ GPU-hours per \emph{billion} candidates. At an
illustrative \$$1.50$/A100-hour rate this is \$$0.67$ vs.\
\$$1.79$ per billion candidate scores — $2.7\times$ fewer
GPU-dollars at any hourly price. The $1.4\!-\!8.9\times$
peak-memory headroom is additional: in a serving stack it converts
to higher request collocation per GPU, not to direct latency.

\begin{table}[t]\centering\small
\caption{Chunked FP16 eager (best chunk size per cell,
oracle-tuned) vs.\ \flash{}, A100-80GB, CUDA-event medians
(legacy warm-cache campaign; frontier cells re-measured under the
full protocol in Tab.~\ref{tab:fwd_chunked}). ``vanilla'' is the
unchunked FP16 eager einsum. The $\times$ column is
chunked$_{\text{best}}$\,/\,\flash{}; note at $B{=}1$K the
\emph{unchunked} vanilla is the faster eager variant (chunking
overhead dominates), and at the textual shape \flash{} is at
parity with it. The best chunk is $256$ at $B{=}1$K and
$1024$ at $B{\geq}5$K for every shape.}
\label{tab:chunked_sweep}
\setlength{\tabcolsep}{3.0pt}
\begin{tabular}{llrrrr}
\toprule
shape & $B$ & vanilla & chunked$_{\text{best}}$ & \flash{} & $\times$ \\
\midrule
textual  & 1K  & $0.19$\,ms  & $0.51$\,ms  & $0.20$\,ms  & $2.6$ \\
long-doc & 1K  & $0.40$\,ms  & $0.53$\,ms  & $0.30$\,ms  & $1.8$ \\
medium   & 1K  & $0.74$\,ms  & $0.88$\,ms  & $0.35$\,ms  & $2.5$ \\
visual   & 1K  & $2.51$\,ms  & $2.65$\,ms  & $1.11$\,ms  & $2.4$ \\
ColPali  & 1K  & $4.26$\,ms  & $4.31$\,ms  & $2.03$\,ms  & $2.1$ \\
visual   & 5K  & $11.6$\,ms  & $11.0$\,ms  & $4.1$\,ms   & $2.7$ \\
visual   & 10K & $23.4$\,ms  & $21.8$\,ms  & $8.4$\,ms   & $2.6$ \\
visual   & 20K & $46.8$\,ms  & $43.9$\,ms  & $16.8$\,ms  & $2.6$ \\
ColPali  & 5K  & $22.6$\,ms  & $21.5$\,ms  & $8.2$\,ms   & $2.6$ \\
ColPali  & 10K & $45.9$\,ms  & $43.0$\,ms  & $16.4$\,ms  & $2.6$ \\
ColPali  & 20K & $92.9$\,ms  & $86.3$\,ms  & $32.6$\,ms  & $2.6$ \\
\bottomrule
\end{tabular}
\end{table}

\section{Chamfer Distance Generalization}
\label{sec:chamfer}
Our inverse-grid CSR backward is operator-agnostic: it applies to any forward
that emits, per source, an integer index into a destination set (\S\ref{sec:backward}).
As a concrete instance, we port the kernel to the \emph{Chamfer distance}
between two point sets $P\in\mathbb{R}^{N\times 3}$ and $Q\in\mathbb{R}^{M\times 3}$,
\[
\mathrm{CD}(P,Q)=\tfrac1N\!\sum_{p}\min_{q}\lVert p-q\rVert^2
              + \tfrac1M\!\sum_{q}\min_{p}\lVert q-p\rVert^2,
\]
the standard 3D point-cloud / shape-matching loss, used since
\citet{chamferorig} as a distance transform and now a default
training signal for point-set models such as
PointNet++~\cite{pointnet2}. It has the same structure as
\maxsim{} with two swaps: a \emph{min}-reduction over the other set
(still idempotent, still rescaler-free) in place of the rowmax,
and squared Euclidean distance in place of the inner product. The naive form materializes the identical
$[N,M]$ pairwise matrix; the fused kernel streams tiles with an online min, and
the backward reuses the saved argmin (nearest-neighbour index) through the
\emph{same} inverse-grid CSR. Measured on A100-80GB against a naive
PyTorch pairwise-distance autograd baseline (\texttt{cuda.synchronize}d
wall-clock), over point sets
$N\!=\!M\!\in\!\{1\text{K},10\text{K},50\text{K},100\text{K}\}$ at
$d\!\in\!\{3,128\}$: gradient cosine $1.00000$ vs.\ naive autograd,
$77\!-\!152\times$ speedup, and $100\mathrm{K}$-point clouds that OOM
the naive form run comfortably. The kernel design transfers beyond
\maxsim{} alone; we do not claim coverage of arbitrary hard-selection operators.

\section{Reproducibility}
\label{sec:repro}
Code, benchmark scripts, and the raw result JSONs that produced every
numbered table in this paper are released at \repourl{}; the
paper-specific artifacts --- the experiment scripts,
\texttt{data/e1\_results/}, and the per-cell audit map
\texttt{PAPER\_NUMBERS.md} (file paths and JSON keys for every
table) --- are pinned on branch \texttt{emnlp2026}. Hardware: NVIDIA A100-SXM4 (40/80\,GB) and
H100 80\,GB HBM3; software: PyTorch 2.8, Triton 3.5/3.6,
\texttt{torch.set\_float32\_matmul\_precision('high')} (TF32 on), CUDA
driver 555.x. \texttt{bash scripts/reproduce\_paper.sh} re-runs the six
core benchmarks (forward latency, HBM traffic, out-of-core, training step,
variable-length, and the corpus-scaling figure source); each remaining
table is reproduced by its own \texttt{benchmarks/bench\_*.py} script
listed alongside the table's row in \texttt{PAPER\_NUMBERS.md}.

\section{Detailed Ablations and Variant Measurements}
\label{sec:details}

For paper-length reasons the per-cell ablation and variant tables sit here rather than in \S\ref{sec:eval}. Each subsection below pairs one table with the one-sentence summary that motivates its content; the surrounding prose in \S\ref{sec:eval} cites these tables by \texttt{\textbackslash ref}.

\subsection{Operator-Design Ablation}\label{app:op_ablation}
Forward fused vs.\ +chunking vs.\ naive einsum on the five canonical shapes; ablation referenced in \S\ref{sec:eval}.

\begin{table}[t]\centering\small
\caption{Forward operator-design ablation on A100-80GB, $N_q{=}1$,
with $B$ scaled down at larger shapes to keep the no-chunk \emph{fused}
variant in budget ($B{=}1024$ textual/long-doc, $512$ medium, $256$
visual, $128$ ColPali; latencies are therefore \emph{not} comparable
across rows or to the $B{=}1$K Tab.~\ref{tab:fwd_abs}). "fused" is
\flash{} with query chunking disabled (\texttt{query\_chunk\_size=None}),
i.e.\ one program per (query, doc) pair, no Q-tile decomposition. "+chunk"
adds the default $128$-token query chunking. Speedup columns are the
naive ms divided by each variant's ms; "$\Delta$ from chunk" is fused $\div$
+chunk and measures the marginal value of the query-chunking choice.
\textbf{Shape note:} the textual row here uses $L_d{=}180$ (short-doc
serving shape, average BEIR-passage tokenised length) whereas
Tab.~\ref{tab:fwd}'s textual row uses $L_d{=}300$ (Pareto-conservative
upper bound); the other four shapes match Tab.~\ref{tab:fwd} in $L_q,L_d,d$ but at smaller $B$, so latencies are not comparable.
The $L_d{=}180$ vs $300$ change explains the textual speedup gap
($2.5\times$ here vs $1.2\times$ in Tab.~\ref{tab:fwd}): the two
rows occupy different kernel-efficiency regimes, so their textual
speedups are not directly comparable.}
\label{tab:op_abl}
\resizebox{\columnwidth}{!}{%
\begin{tabular}{lcccccc}
\toprule
Shape ($L_q,L_d$) & naive ms & fused ms & +chunk ms & naive$\div$fused & naive$\div$+chunk & $\Delta$ from chunk \\
\midrule
textual ($32,180$)   & $0.30$ & $0.13$ & $0.13$ & $2.4\times$ & $\mathbf{2.5\times}$ & $1.02\times$ \\
long-doc ($32,1024$) & $1.23$ & $0.25$ & $0.25$ & $4.9\times$ & $\mathbf{5.0\times}$ & $1.01\times$ \\
medium ($128,1024$)  & $0.89$ & $0.23$ & $0.22$ & $3.9\times$ & $\mathbf{4.1\times}$ & $1.05\times$ \\
visual ($512,1024$)  & $1.03$ & $0.84$ & $0.36$ & $1.2\times$ & $\mathbf{2.9\times}$ & $\mathbf{2.4\times}$ \\
ColPali ($1024^2$)   & $0.94$ & $0.76$ & $0.39$ & $1.2\times$ & $\mathbf{2.4\times}$ & $\mathbf{1.9\times}$ \\
\bottomrule
\end{tabular}}
\par\smallskip
\footnotesize Same timing protocol as Tab.~\ref{tab:fwd}: matched-precision
naive (FP16 inputs, FP32 accumulation), FP16$\to$FP32 cast hoisted out of every
timed region, $50$ post-warmup CUDA-event medians (App.~\ref{app:setup}).
\end{table}

\subsection{Multi-Query Serving Sweep}\label{app:multiquery}
$N_q$ concurrent queries batched against a single corpus, fixed
$B$ per shape; the production rerank-service operating point where
requests are buffered for a few ms before a single GPU dispatch.
A100-80GB, $5$ warmup + $30$ timed CUDA-event measurements per
cell; median latency, derived queries-per-second, peak memory.
\flash{} scales near-linearly in $N_q$ with much higher absolute
throughput and a $2.1\!-\!4.7\times$ per-call lead over naive at
every cell that naive fits.

\begin{table}[h]\centering\small\setlength{\tabcolsep}{4pt}
\caption{Multi-query serving: \flash{} vs a \emph{textbook} FP32
naive (full-FP32 multiply, dtype cast inside the timed region) at
three shapes, A100-80GB. This naive is slower than the
matched-precision naive of Tab.~\ref{tab:fwd}/\ref{tab:fwd_abs}, so
the per-call speedups here are larger and \emph{not} comparable across
tables. Latency is median per-call ms; QPS is $N_q$/latency.}
\label{tab:multiquery}
\begin{tabular}{lrrrrr}
\toprule
Shape ($B$, $L_q, L_d$) & $N_q$ & \flash{} ms & QPS & naive ms & speedup \\
\midrule
\multirow{5}{*}{textual ($1024,32,180$)}
 & $1$  & $0.13$ & $7644$  & $0.27$  & $2.1\times$ \\
 & $4$  & $0.18$ & $21941$ & $0.43$  & $2.3\times$ \\
 & $8$  & $0.23$ & $34412$ & $0.64$  & $2.8\times$ \\
 & $16$ & $0.42$ & $37753$ & $1.29$  & $3.0\times$ \\
 & $32$ & $0.72$ & $\mathbf{44729}$ & $2.05$  & $2.9\times$ \\
\midrule
\multirow{5}{*}{long-doc ($1024,32,1024$)}
 & $1$  & $0.27$ & $3742$  & $1.25$  & $4.7\times$ \\
 & $4$  & $0.44$ & $9131$  & $1.75$  & $4.0\times$ \\
 & $8$  & $0.71$ & $11192$ & $2.50$  & $3.5\times$ \\
 & $16$ & $1.28$ & $12505$ & $3.93$  & $3.1\times$ \\
 & $32$ & $2.57$ & $\mathbf{12459}$ & $7.45$  & $2.9\times$ \\
\midrule
\multirow{5}{*}{ColPali ($128,1024,1024$)}
 & $1$  & $0.34$ & $2942$  & $0.94$  & $2.8\times$ \\
 & $4$  & $0.92$ & $4339$  & $3.39$  & $3.7\times$ \\
 & $8$  & $1.70$ & $4700$  & $6.95$  & $4.1\times$ \\
 & $16$ & $3.29$ & $4860$  & $14.34$ & $4.4\times$ \\
 & $32$ & $6.39$ & $\mathbf{5010}$ & $28.60$ & $\mathbf{4.5\times}$ \\
\bottomrule
\end{tabular}
\end{table}

\subsection{Out-of-Core Corpus Scoring}
Streaming a host-resident corpus through the GPU at fixed peak memory; referenced in \S\ref{sec:eval}.

\begin{table}[t]\centering\small
\caption{Out-of-core scoring (ColPali $L_q{=}L_d{=}1024$, A100-80GB): one query
vs.\ a host-resident corpus streamed to the GPU in $20\mathrm{K}$-document blocks.
GPU peak is flat regardless of corpus size; the last two corpora exceed the
card's $80$\,GiB (${\approx}86$\,GB) capacity.}
\label{tab:ooc}
\begin{tabular}{rccc}
\toprule
corpus $B$ & embeddings & exceeds VRAM & GPU peak \\
\midrule
$100$K & $26$\,GB  & no  & $5.5$\,GB \\
$250$K & $66$\,GB  & no  & $5.2$\,GB \\
$500$K & $131$\,GB & yes & $5.3$\,GB \\
$750$K & $197$\,GB & yes & $5.3$\,GB \\
\bottomrule
\end{tabular}
\end{table}

\subsection{Backward-Path Ablation}\label{app:bwd_ablations}
CSR / plain atomic / atomic-unified at the same six contrastive training shapes on H100; the dispatcher default is \emph{atomic-unified} (\S\ref{sec:backward}). The CSR cost-decomposition subsection below pairs with this one: build is $54\!-\!78\%$ of the backward, which is why the unified path (no CSR build) matches CSR at large shapes.

\begin{table}[t]\centering\small
\caption{Backward-path ablation on H100, contrastive in-batch
negatives. Three backward paths at the same forward shape: inverse-
grid CSR, plain atomic scatter, and the fused unified backward (the
dispatcher default). Bold marks the fastest path on each row.
Memory is the step peak.}
\label{tab:bwd_abl}
\resizebox{\columnwidth}{!}{%
\begin{tabular}{lcccccc}
\toprule
Shape & CSR ms & CSR peak & atomic ms & atomic peak & unified ms & unified peak \\
\midrule
ColBERT $B{=}128$  ($L_q{=}32$)   & $0.96$  & $0.04$\,GB & $3.95$   & $0.03$\,GB & $\mathbf{0.76}$ & $0.03$\,GB \\
ColBERT $B{=}512$  ($L_q{=}32$)   & $\mathbf{5.76}$ & $0.43$\,GB & $18.62$  & $0.13$\,GB & $6.21$ & $0.13$\,GB \\
ColBERT $B{=}1024$ ($L_q{=}32$)   & $\mathbf{21.21}$ & $1.56$\,GB & $50.85$ & $0.33$\,GB & $23.53$ & $0.33$\,GB \\
ColPali $B{=}32$   ($L_q{=}1024$) & $1.87$  & $0.09$\,GB & $28.48$  & $0.06$\,GB & $\mathbf{1.86}$ & $0.06$\,GB \\
ColPali $B{=}64$   ($L_q{=}1024$) & $\mathbf{5.81}$ & $0.27$\,GB & $58.13$ & $0.12$\,GB & $6.05$ & $0.12$\,GB \\
ColPali $B{=}128$  ($L_q{=}1024$) & $\mathbf{21.02}$ & $0.88$\,GB & $136.86$ & $0.28$\,GB & $22.65$ & $0.28$\,GB \\
\bottomrule
\end{tabular}}
\end{table}

\subsection{Backward Component Breakdown}
Per-stage timing of the CSR backward on A100-80GB for a \emph{single}
query ($N_q{=}1$) against $B$ documents, isolating the per-call kernel
costs. These are per-query kernel times, \emph{not} the $N_q{=}B$
in-batch-negatives step of Tab.~\ref{tab:train} or the ablation of
Tab.~\ref{tab:bwd_abl} (whose totals are far larger); what transfers
across $N_q$ is the component \emph{split}, which shows why the fused
unified path (skipping the CSR build) catches up at large shapes.

\begin{table}[t]\centering\small
\caption{Backward component breakdown for the CSR (inverse-grid) training
path on A100-80GB, at the same shapes as Table~\ref{tab:bwd_abl} but for a single query ($N_q{=}1$). We time the
forward and each backward stage in isolation with CUDA events
(independent medians); ``bwd pieces'' is a \emph{separate} timing of
the three backward stages executed together, so it need not equal the
sum of the per-stage medians; ``autograd step'' is one
\texttt{loss.backward()} round-trip for cross-validation. Parenthesised
percentages are each stage's share of the backward total.}
\label{tab:bwd_breakdown}
\resizebox{\columnwidth}{!}{%
\begin{tabular}{lcccccc}
\toprule
Shape & fwd ms & CSR build ms & $\nabla Q$ ms & $\nabla D$ ms & bwd pieces ms & autograd step ms \\
\midrule
ColBERT $B{=}128$  ($L_q{=}32$)    & $0.19$ & $0.45$ ($76\%$) & $0.07$ ($12\%$) & $0.08$ ($13\%$) & $0.58$  & $0.55$ \\
ColBERT $B{=}512$  ($L_q{=}32$)    & $0.20$ & $0.49$ ($66\%$) & $0.09$ ($12\%$) & $0.13$ ($17\%$) & $0.75$  & $0.63$ \\
ColBERT $B{=}1024$ ($L_q{=}32$)    & $0.20$ & $0.55$ ($54\%$) & $0.15$ ($15\%$) & $0.23$ ($23\%$) & $1.02$  & $0.76$ \\
ColPali  $B{=}32$   ($L_q{=}1024$) & $0.46$ & $0.51$ ($78\%$) & $0.07$ ($10\%$) & $0.08$ ($13\%$) & $0.65$  & $1.33$ \\
ColPali  $B{=}64$   ($L_q{=}1024$) & $0.46$ & $0.52$ ($75\%$) & $0.07$ ($10\%$) & $0.12$ ($18\%$) & $0.70$  & $1.16$ \\
ColPali  $B{=}128$  ($L_q{=}1024$) & $0.80$ & $0.53$ ($67\%$) & $0.08$ ($10\%$) & $0.19$ ($24\%$) & $0.79$  & $1.74$ \\
\bottomrule
\end{tabular}}
\par\smallskip
\footnotesize Two observations. (i) CSR build is the dominant fixed cost
because it does the global bincount + prefix-sum that lets the two
$\nabla$ kernels read row-owned data without atomics; the $\nabla$ kernels
together are a fraction of the build cost at every tested shape.
(ii) The autograd-roundtrip total drifts above the pieces sum at ColPali
shapes because the autograd backward computes both $\nabla Q$ and a full
FP32 $\nabla D$ allocation/cast pass not charged to the per-kernel
measurements.
\end{table}

\paragraph{Alternative: chunked recompute.} Gradient checkpointing
(the \emph{recompute} rows of Tab.~\ref{tab:train}) is the other way
to fit $B{=}128$ in-batch negatives where naive autograd OOMs. It
recovers feasibility but at $8\times$ \flash{}'s step time and
${\sim}45\times$ its peak: it re-pays the forward matmul on the
backward and still materialises each $[B_{\mathrm{blk}},L_q,L_d]$
block, so the materialised-tensor cost returns per block
(\texttt{bench\_recompute\_train\_*.json}).

\subsection{Variable-Length Scoring}\label{sec:varlen}
\texttt{cu\_seqlens} vs.\ padded baseline by document-length fill ratio; the up-to-$4.6\times$ number quoted in \S\ref{sec:eval} comes from the highly-ragged row.

\begin{table}[t]\centering\small
\caption{Variable-length scoring: \flash{} (\texttt{cu\_seqlens}) vs.\ naive
padded, by document-length distribution ($L_d^{\max}{=}512$). The win tracks the
fill ratio $f_{\mathrm{fill}}=\sum_b L_d^{(b)}/(B L_d^{\max})$ (higher fill, less padding to skip, less
speedup); $f_{\mathrm{fill}}$ is distinct from the ranking $\rho$ used elsewhere.}
\label{tab:varlen}
\begin{tabular}{lccccc}
\toprule
Length distribution & $f_{\mathrm{fill}}$ & avg $L_d$ & naive ms & \flash{} ms & speedup \\
\midrule
uniform $[256,512]$ & $0.76$ & $390$ & $0.553$ & $0.171$ & $3.24\times$ \\
HotpotQA-like       & $0.24$ & $120$ & $0.555$ & $0.130$ & $4.27\times$ \\
highly ragged       & $0.14$ & $\phantom{0}71$ & $0.553$ & $0.120$ & $\mathbf{4.59\times}$ \\
\bottomrule
\end{tabular}
\par\smallskip\footnotesize $B{=}1000$ documents, $L_d^{\max}{=}512$, $L_q{=}32$, $d{=}128$ on A100-80GB.
\end{table}

\subsection{INT8 across Shapes}
INT8 vs.\ FP16-\flash{}, dequant-then-naive, and naive einsum at five shapes; precision metrics on the right.

\begin{table}[t]\centering\small
\caption{INT8$\times$INT8 across the five canonical shapes
(fixed $B$, see column~$B$; the textual row here uses the short-doc
serving length $L_d{=}180$, vs.\ Tab.~\ref{tab:fwd}'s $L_d{=}300$),
single A100-80GB. "naive" is the textbook
FP32 \texttt{einsum}+\texttt{max}+\texttt{sum}; "dequant" is the
dequantize-then-naive baseline; "FP16 \flash{}" is the FP16 forward
kernel of Tab.~\ref{tab:fwd}. Last three columns are the speedup of
INT8 \flash{} over each baseline (higher is better for INT8).
Storage column is the $D$-tensor footprint ratio.}
\label{tab:int8}
\resizebox{\columnwidth}{!}{%
\begin{tabular}{lcccccccc}
\toprule
Shape & $B$ & FP16 \flash{} & INT8 \flash{} & vs naive & vs dequant & vs FP16 \flash{} & Spearman $\rho$ / top-$20$ & $D$ storage \\
\midrule
textual ($32,180$)   & $1024$ & $0.14$ & $0.15$ & $2.1\times$ & --- & $0.93\times$ & $0.9997$ / $\mathbf{100\%}$ & $1.97\times$ \\
long-doc ($32,1024$) & $1024$ & $0.26$ & $0.21$ & $\mathbf{6.0\times}$ & --- & $1.27\times$ & $0.9995$ / $95\%$            & $1.97\times$ \\
medium ($128,1024$)  & $512$  & $0.23$ & $0.24$ & $3.7\times$ & --- & $0.98\times$ & $0.9996$ / $\mathbf{100\%}$ & $1.97\times$ \\
visual ($512,1024$)  & $256$  & $0.35$ & $0.34$ & $3.0\times$ & --- & $1.04\times$ & $0.9992$ / $\mathbf{100\%}$ & $1.97\times$ \\
ColPali ($1024^2$)   & $128$  & $0.35$ & $0.30$ & $3.0\times$ & $3.3\times$ & $1.07\times$ & $0.9993$ / $\mathbf{100\%}$ & $1.97\times$ \\
\bottomrule
\end{tabular}}
\par\smallskip
\footnotesize "vs dequant" is only reported at ColPali because that is the
working point for the original "dequantize-then-naive" baseline in industry
practice; at smaller shapes the dequant baseline is dominated by the
allocation cost of the materialised FP32 $D$ tensor and is a strict
multiple of "vs naive". See Tab.~\ref{tab:int8_B} for the dequant-baseline
sweep at ColPali shape.
\end{table}

\subsection{INT8 $B$-Sweep at ColPali}
INT8 latency and ranking fidelity as corpus size $B$ grows on ColPali shape.

\begin{table}[t]\centering\small
\caption{INT8$\times$INT8 at ColPali shape ($L_q{=}L_d{=}1024$, $d{=}128$),
sweeping the corpus size $B$ from $128$ to $10$K on A100-80GB. INT8's win
over both naive baselines grows monotonically with $B$ as the corpus
exceeds the GPU's L2 cache and memory bandwidth becomes the bottleneck
(L2 on A100 is $40$\,MB; the $D$ tensor exceeds it at $B\!\gtrsim\!160$
FP16). The win over FP16 \flash{} also grows but more gently, the FP16
forward is already streaming-light.}
\label{tab:int8_B}
\resizebox{\columnwidth}{!}{%
\begin{tabular}{rcccccc}
\toprule
$B$ & naive ms & dequant ms & FP16 ms & INT8 ms & INT8$\div$naive & INT8$\div$dequant \\
\midrule
$128$   & $0.94$  & $1.03$  & $0.34$  & $0.31$  & $3.0\times$ & $3.3\times$ \\
$256$   & $1.84$  & $2.02$  & $0.57$  & $0.48$  & $3.8\times$ & $4.2\times$ \\
$512$   & $3.62$  & $3.95$  & $0.98$  & $0.80$  & $4.5\times$ & $4.9\times$ \\
$1024$  & $7.34$  & $8.01$  & $1.75$  & $1.45$  & $5.0\times$ & $5.5\times$ \\
$2048$  & $14.83$ & $16.13$ & $3.40$  & $2.75$  & $5.4\times$ & $5.9\times$ \\
$4096$  & $29.87$ & $32.28$ & $6.71$  & $5.37$  & $5.6\times$ & $6.0\times$ \\
$10000$ & $74.35$ & $80.73$ & $16.19$ & $12.77$ & $\mathbf{5.8\times}$ & $\mathbf{6.3\times}$ \\
\bottomrule
\end{tabular}}
\par\smallskip
\footnotesize "dequant" is the dequantize-then-naive baseline: materialise
the FP32 $D$ tensor from the INT8 store, then \texttt{einsum}+\texttt{max}+\texttt{sum}.
INT8 vs.\ FP16 \flash{} stays a modest $1.07\times$ at $B{=}128$ rising
to $1.27\times$ at $B{=}10$K (the kernel's win over FP16 is bounded by the
ratio of dequant-and-multiply cost to streaming-D cost); INT8's larger
$3\!-\!6\times$ wins are over the naive FP32 and dequant-then-FP32 baselines
that some production rerankers still run.
\end{table}

\subsection{Fat-Embedding Latency Cliff and Split-d Forward}
\label{sec:splitd}
At $d > 512$ the in-one-shot forward's operand tile exceeds the per-SM
SRAM budget (A100 $164$\,KB; H100 $228$\,KB), Triton spills to local
memory, and latency rises to ${\approx}1.4\!-\!4.7\times$ (H100) /
${\approx}2.2\!-\!19.3\times$ (A100) above a linear extrapolation
from $d{=}512$, reaching $17.3$ / $99.7$\,ms at $d{=}2048$
(Tab.~\ref{tab:fat_emb_cliff}). The split-d forward
(\texttt{flash\_maxsim\_splitd}) adds an inner $\text{BLOCK\_K}$ loop
that tiles the embedding dim and accumulates in FP32 across $d$-tiles;
the reduction structure is unchanged, so it is a drop-in replacement
gated on $d > 512$ in the dispatcher (per-arch heuristic launch-config
table picks \texttt{(BLOCK\_Q,BLOCK\_D,BLOCK\_K,num\_warps,num\_stages)}
per $d$-bucket per arch). It gives $\mathbf{2\!-\!20\times}$ over the
spilling in-one-shot path at $d \in \{768,1024,2048\}$ with identical
correctness (max $|\Delta|$ vs.\ FP32 in $[\,10^{-5},10^{-4}\,]$;
Tab.~\ref{tab:splitd_perf}).

\begin{table}[t]\centering\small
\caption{In-one-shot forward latency vs.\ embedding dim $d$ on the
ColPali shape ($B{=}128$, $L_q{=}L_d{=}1024$). The cliff at $d > 512$
on both arches reflects the operand tile exceeding the per-SM SRAM
budget and Triton spilling to local memory. Correctness (max $|\Delta|$
vs.\ FP32) is preserved across the sweep; only latency degrades.}
\label{tab:fat_emb_cliff}
\resizebox{\columnwidth}{!}{%
\begin{tabular}{rcccc}
\toprule
$d$ & in-one-shot A100 ms & in-one-shot H100 ms & approx.\ tile (KB) & ratio vs.\ d$=$512 \\
\midrule
$128$  & $0.36$  & $0.36$  & $32$   & $0.27\times$  \\
$256$  & $0.59$  & $0.55$  & $64$   & $0.45\times$  \\
$384$  & $1.28$  & $0.96$  & $96$   & $1.00\times$  \\
$512$  & $1.29$  & $0.93$  & $128$  & $1.00\times$  \\
$768$  & $5.19$  & $2.68$  & $192$  & $\mathbf{4.0\times}$  \\
$1024$ & $5.59$  & $2.58$  & $256$  & $\mathbf{4.3\times}$  \\
$2048$ & $99.7$  & $17.3$  & $512$  & $\mathbf{77\times}$ (A100) \\
\bottomrule
\end{tabular}}
\end{table}

\subsection{Split-d Speedup}
Split-d forward vs.\ in-one-shot at $d > 512$ on the same ColPali shape; speedups are over the spilling baseline.

\begin{table}[t]\centering\small
\caption{Split-d forward vs.\ in-one-shot on the same ColPali shape
($B{=}128$, $L_q{=}L_d{=}1024$). Split-d is dispatched at $d > 512$
in the public API; below the threshold the in-one-shot kernel is
faster (no inner-loop overhead) and remains the default. Correctness:
max $|\Delta|$ vs.\ FP32 stays in $[\,10^{-5},10^{-4}\,]$ at every
$d$, matching the in-one-shot kernel's FP16-cast noise.}
\label{tab:splitd_perf}
\resizebox{\columnwidth}{!}{%
\begin{tabular}{rcccc}
\toprule
$d$ & A100 in-shot $\to$ split-d (ms) & speedup & H100 in-shot $\to$ split-d (ms) & speedup \\
\midrule
$768$  & $5.19 \to \mathbf{2.43}$  & $\mathbf{2.13\times}$ & $2.68 \to \mathbf{0.76}$  & $\mathbf{3.53\times}$ \\
$1024$ & $5.59 \to \mathbf{2.60}$  & $\mathbf{2.15\times}$ & $2.58 \to \mathbf{1.07}$  & $\mathbf{2.41\times}$ \\
$2048$ & $99.7 \to \mathbf{4.91}$  & $\mathbf{20.3\times}$ & $17.3 \to \mathbf{2.06}$  & $\mathbf{8.40\times}$ \\
\bottomrule
\end{tabular}}
\par\smallskip
\footnotesize Removes the in-one-shot kernel's effective $d$ ceiling
for any \maxsim{} workload at $d > 512$. The algorithmic SRAM budget
scales with $\text{BLOCK\_K}$ rather than $d$, so the kernel is
expected to scale to larger $d$ as well, but we measure only up to
$d{=}2048$. Training through the autograd Function (forward via
split-d, backward via the existing CSR or atomic-unified path) is
also verified end-to-end at every $d \in \{512,768,1024,2048\}$:
$\nabla Q$ and $\nabla D$ match a naive PyTorch autograd reference
to FP16-backward noise (max $|\Delta|\!\leq\!10^{-3}$), so
fat-embedding training works without a separate backward kernel
— one $d$-vector per backward program is register-friendly to
$d{\approx}4096$.
\end{table}

\section{Experimental Setup Details}\label{app:setup}
The full benchmarking protocol distilled from this paper's
measurement corrections — precision tiers, compile flavours,
fresh-process memory, interleaved timing, dispersion scope — is in
App.~\ref{app:bench_protocol}.
The five canonical workload shapes used throughout \S\ref{sec:eval}
(Tab.~\ref{tab:shapes}), precision protocols per backend
(Tab.~\ref{tab:precision}), the baseline glossary
(Tab.~\ref{tab:baselines}), the HBM-traffic table behind
Theorem~\ref{thm:io} (Tab.~\ref{tab:traffic}), and the per-$B$ peak
memory (Tab.~\ref{tab:mem}) are collected here. $L_q$ and $L_d$ are
encoder maximum token / patch counts (we use the maximum, not
per-instance, for reproducible latency comparisons).

\begin{table}[h]\centering\small
\caption{Notation used throughout the paper. The workload shapes
(\textit{textual}, \textit{long-doc}, \textit{medium},
\textit{visual}, \textit{ColPali}) are defined in
Tab.~\ref{tab:shapes}.}
\label{tab:notation}
\begin{tabular}{@{}l p{0.78\linewidth}@{}}
\toprule
Symbol & Meaning \\
\midrule
$N_q$ & queries scored in one call ($N_q{=}1$: single-query rerank; arbitrary $N_q$: batched multi-query serving; typically $N_q{=}B$: in-batch-negatives training) \\
$B$   & number of documents (candidate-set size at inference, batch size in training) \\
$L_q$ & query token / patch count (encoder maximum) \\
$L_d$ & document token / patch count (encoder maximum) \\
$d$   & per-token embedding dimension ($128$ unless noted) \\
$S$   & the materialised $[N_q,B,L_q,L_d]$ query$\times$document similarity tensor \\
$\rho$ & Spearman rank correlation of scores to the FP32 reference \\
\bottomrule
\end{tabular}
\end{table}

\begin{table}[h]\centering\small
\caption{Workload shapes used throughout \S\ref{sec:eval} and their
realised encoder/dataset provenance. $L_q$ and $L_d$ are the
encoder's maximum token or patch counts; $d$ is the embedding
dimensionality. Text-query$\to$page ColPali retrieval has a short
text $L_q$ and corresponds to the long-doc / medium rows; the
visual and ColPali rows are the page-as-query (image-to-image)
regime where $L_q\!=\!L_d$.}
\label{tab:shapes}
\begin{tabular}{@{}l c c c p{0.58\linewidth}@{}}
\toprule
shape & $L_q$ & $L_d$ & $d$ & encoder / corpus \\
\midrule
textual   & $32$    & $300$  & $128$ & ColBERT \cite{colbert} on BEIR short-text (e.g.\ arguana, scidocs) \\
long-doc  & $32$    & $1024$ & $128$ & ColBERT on long-document corpora (LoCoV1, HotpotQA full passages) \\
medium    & $128$   & $1024$ & $128$ & ColBERT-long / Jina-ColBERT-v2 at $L_q\!\approx\!128$ training-time queries \\
visual    & $512$   & $1024$ & $128$ & ColPali \cite{colpali} half-res pages, page-as-query ($\approx\!512$ patches each) \\
ColPali   & $1024$  & $1024$ & $128$ & ColPali full-res pages, page-as-query ($1024$ patches as both query and corpus; visual doc-to-doc / dedup) \\
\bottomrule
\end{tabular}
\end{table}

\begin{table}[h]\centering\small
\caption{Precision protocols for each method. The matched naive path
and \flash{} both accumulate in FP32; the multiply unit is TF32
tensor cores for the matched naive and FP16/BF16 tensor cores for
\flash{}, matching what \texttt{compile-MA} typically selects. For
the matched naive path, ``FP16'' input denotes the stored embedding
dtype: inputs are pre-cast to FP32 outside the timed region, and the
einsum then runs on TF32 tensor cores. The
two \emph{naive in-dtype} rows accumulate in the input dtype and are
reference points for the bf16 sensitivity analysis in
\S\ref{sec:correct} (naive bf16 loses precision catastrophically at
ColPali scale; \flash{} does not).}
\label{tab:precision}
\begin{tabular}{lcccc}
\toprule
Path & Input & Multiply & Accum.\ & Output \\
\midrule
naive eager (matched)         & FP16 & TF32 TC & FP32 & FP32 \\
naive eager (pure FP16)       & FP16 & FP16 TC & FP16 & FP32 \\
naive eager (bf16, in-dtype)  & BF16 & BF16 TC & BF16 & FP32 \\
\texttt{compile-MA}           & FP16 & varies  & FP32 & FP32 \\
\flash{} (forward, FP16)      & FP16 & FP16 TC & FP32 & FP32 \\
\flash{} (forward, BF16)      & BF16 & BF16 TC & FP32 & FP32 \\
FP32 reference (TF32 off)     & FP32 & FP32    & FP32 & FP32 \\
\bottomrule
\end{tabular}
\end{table}

\begin{table}[h]\centering\small
\caption{Baselines used in \S\ref{sec:eval}. Each row is a backend we
time, what it is, and the question it isolates.}
\label{tab:baselines}
\begin{tabular}{p{3.5cm} p{5.5cm} p{5.5cm}}
\toprule
Name & Backend & Question it answers \\
\midrule
naive eager (matched) & PyTorch \texttt{einsum + max + sum}, FP16 in / FP32 accum / TF32 TC & ``How much does fusion save vs PyTorch at fair precision?'' \\
chunked FP16 eager & corpus-chunked FP16 \texttt{einsum + max + sum}, oracle chunk size (App.~\ref{app:chunked_eager}) & ``How strong is the deployed memory workaround?'' \\
\texttt{compile-MA}    & \texttt{torch.compile(max-autotune)} of the same expression                & ``Can Inductor already fuse this?'' \\
chunked + \texttt{compile-MA} & manual corpus chunking + \texttt{compile-MA} (App.~\ref{sec:compile}) & ``Does manual chunking close the memory gap?'' \\
dequant + naive       & dequantise INT8 D to FP32, then naive einsum (production INT8 path)        & ``What does INT8 buy over the standard dequant pipeline?'' \\
FP32 reference        & FP32 inputs / FP32 accum / TF32 \emph{off}                                 & ``How close is the kernel to true FP32?'' (correctness only) \\
naive eager (FP16 in-dtype)  & FP16 in / FP16 accum                                              & precision diagnostic for \S\ref{sec:correct} only \\
naive eager (bf16 in-dtype)  & BF16 in / BF16 accum                                              & precision diagnostic for the bf16 silent-failure analysis \\
\bottomrule
\end{tabular}
\end{table}

\begin{table}[h]\centering\small\setlength{\tabcolsep}{4pt}
\caption{Absolute forward latency (ms) and peak memory (GB) per
shape at $B{=}1$K, A100-80GB — same campaign as Tab.~\ref{tab:fwd}
(full protocol, App.~\ref{app:bench_protocol}). Compile peaks are
omitted (graph-pool reservations are not comparable to allocator
peaks). Bold = best per column group.}
\label{tab:fwd_abs}
\setlength{\tabcolsep}{3.5pt}
\begin{tabular}{lrrrrrr}
\toprule
 & \multicolumn{4}{c}{latency (ms)} & \multicolumn{2}{c}{peak (GB)} \\
\cmidrule(lr){2-5} \cmidrule(lr){6-7}
Shape & FP32 & MA & nocg & \flash{} & FP32 & \flash{} \\
\midrule
textual   & $0.24$ & $0.47$ & $0.20$ & $\mathbf{0.20}$ & $0.38$ & $\mathbf{0.34}$ \\
long-doc  & $0.60$ & $1.31$ & $0.37$ & $\mathbf{0.31}$ & $1.03$ & $\mathbf{0.90}$ \\
medium    & $1.07$ & $1.77$ & $0.66$ & $\mathbf{0.38}$ & $1.43$ & $\mathbf{0.90}$ \\
visual    & $3.32$ & $3.85$ & $1.79$ & $\mathbf{0.98}$ & $3.00$ & $\mathbf{0.90}$ \\
ColPali   & $6.63$ & $6.82$ & $3.44$ & $\mathbf{1.77}$ & $5.12$ & $\mathbf{0.90}$ \\
\bottomrule
\end{tabular}
\end{table}

\begin{table}[h]\centering\small
\caption{HBM traffic at $B{=}1$K: bytes the algorithm reads and writes.
\flash{}'s $0.26$\,GB is the dominant corpus-side document-read floor
at $L_d{=}1024, d{=}128$ in the single-query rerank setting, where $Q$
is loaded once and reused across the corpus; it grows linearly with
$B L_d d$. (For page-as-query, $L_q{=}L_d$, the one-time query read
matches one document; Theorem~\ref{thm:io}'s per-pair
$\Theta(N_q B (L_q{+}L_d) d)$ is the conservative bound when $Q$ is
instead streamed per document.) The naive path's traffic scales with
$B L_q L_d$ (the materialised $S$ tensor), which is what
Theorem~\ref{thm:io} eliminates.}
\label{tab:traffic}
\begin{tabular}{lccc}
\toprule
Shape ($L_q,L_d$) & naive HBM & \flash{} HBM & ratio \\
\midrule
medium ($128,1024$)  & $1.31$\,GB & $0.26$\,GB & $5\times$ \\
visual ($512,1024$)  & $4.46$\,GB & $0.26$\,GB & $17\times$ \\
ColPali ($1024^2$)   & $8.65$\,GB & $0.26$\,GB & $33\times$ \\
\bottomrule
\end{tabular}
\end{table}

\begin{table}[h]\centering\small
\caption{ColPali corpus scaling ($L_q{=}L_d{=}1024$): peak memory.
Naive shown on A100-40GB (FP16 intermediate) and A100-80GB
(matched FP32); \flash{} peak tracks the document embeddings. Both
naive paths OOM by $B{=}20$K (latency in Fig.~\ref{fig:scaling}).}
\label{tab:mem}
\begin{tabular}{lccc}
\toprule
$B$ & naive 40GB (FP16) & naive 80GB (FP32) & \flash{} \\
\midrule
$10$K  & $23.9$\,GB & $47.2$\,GB & $2.9$\,GB\,$^{*}$ \\
$20$K  & \textbf{OOM} & \textbf{OOM} & $5.2$\,GB \\
$50$K  & OOM & OOM & $13.1$\,GB \\
$100$K & OOM & OOM & $26.2$\,GB \\
\bottomrule
\end{tabular}
\par\smallskip
\footnotesize $^{*}$Raw JSON: $2.63$\,GB $D$-footprint alone
($10^4\!\cdot\!1024\!\cdot\!128\!\cdot\!2$); the $2.9$\,GB includes
the one-time Triton autotune scratch buffer (allocated once at first
call, reused thereafter — steady-state peak in a long inference loop
is closer to $2.63$\,GB).
\end{table}

\paragraph{Fairness of precision and timing comparisons.}
Each cell in \S\ref{sec:eval} reports the same workload run under
each compared backend on the same machine, in the same process, in
the same precision protocol of Tab.~\ref{tab:precision}: FP16 inputs
with FP32 accumulation on both \flash{} and the matched naive
baseline (TF32 tensor-core matmul for naive; FP16 tensor-core matmul
for \flash{}; both produce FP32 outputs). The FP32 reference used
for correctness checks runs in true FP32 with TF32 explicitly
disabled (\texttt{set\_float32\_matmul\_precision('highest')}).
\texttt{compile-MA} is invoked with the same matched-precision
expression as the eager naive baseline and uses Inductor's default
accumulation choice (FP32 in our measurements). All FP16$\to$FP32
casts are hoisted out of every timed region — the cast cost is
charged to no method. Allocation, autotune trials, and JIT-compile
cost likewise sit outside the timed region; every cell is steady-state
kernel cost measured over $50$ post-warmup runs.

\section{Backward Memory Footprint}\label{app:bwd_mem}
To make the memory advantage of the fused backward concrete, we trace
the largest transient that each implementation allocates between
forward and backward at ColPali in-batch-negatives shape
($N_q{=}B$, $L_q{=}L_d{=}1024$, $d{=}128$, FP16 inputs / FP32
accumulation). Two facts dominate. \textbf{(i)} The vanilla autograd
backward does not just retain the forward intermediate $S$; it
allocates a second tensor of the same shape, $\nabla S$, to scatter
the upstream gradient through the chain rule of $\max$ and
$\mathrm{sum}$. The $[N_q,B,L_q,L_d]$ shape thus appears
\emph{twice} per step (the FP32 cast for matched accumulation
doubles each one again, into $\nabla S \in \text{FP32}$). \textbf{(ii)}
The \flash{} backward saves only the integer argmax produced by the
forward (one $L_q$-vector of \texttt{int32} per $(q,d)$ pair); every
gradient it needs is reconstructed from $Q$, $D$, and that index
buffer on the fly. Tab.~\ref{tab:bwd_mem} works through the bytes;
the fused atomic-unified backward (the default) drops the CSR build
too — only the argmax index buffer survives between passes.

\begin{table}[h]\centering\small
\caption{Backward memory footprint at ColPali in-batch-negatives
($N_q{=}B$, $L_q{=}L_d{=}1024$, $d{=}128$, FP16 inputs / FP32
accumulation). Rows show what each backward path allocates between
forward and backward, beyond the always-present inputs and parameter
grads. "Vanilla" is PyTorch autograd over the textbook
\texttt{einsum + max + sum} reduction; "\flash{} (CSR)" is the
inverse-grid path of Alg.~\ref{alg:bwd}; "\flash{} (unified)" is the
fused dQ+dD kernel used by default. $S, \nabla S \in \mathbb{R}^{N_q
\times B \times L_q \times L_d}$ FP32. Totals are an
\emph{idealised floor}: they exclude the always-present $Q$, $D$,
their autograd gradients, workspace, and the $[N_q,B]$ score matrix
(measured decomposition: Tab.~\ref{tab:mem_audit}).}
\label{tab:bwd_mem}
\begin{tabular}{l l r r r r}
\toprule
$B$ & Path & $S$ (fwd) & $\nabla S$ (bwd) & argmax + CSR & total transient \\
\midrule
\multirow{3}{*}{$64$}
 & vanilla autograd  & $17.2$\,GB & $17.2$\,GB & ---            & $34.4$\,GB \\
 & \flash{} (CSR)    & ---        & ---        & $17$\,MB + $195$\,MB & $\mathbf{0.21}$\,GB \\
 & \flash{} (unified) & ---       & ---        & $17$\,MB      & $\mathbf{0.02}$\,GB \\
\midrule
\multirow{3}{*}{$128$}
 & vanilla autograd  & $68.7$\,GB & $68.7$\,GB & ---            & $137.4$\,GB (OOMs the card) \\
 & \flash{} (CSR)    & ---        & ---        & $67$\,MB + $762$\,MB & $\mathbf{0.83}$\,GB \\
 & \flash{} (unified) & ---       & ---        & $67$\,MB      & $\mathbf{0.07}$\,GB \\
\bottomrule
\end{tabular}
\par\smallskip
\footnotesize $|S| = N_q \cdot B \cdot L_q \cdot L_d \cdot 4$\,B in
FP32 (decimal GB throughout): $64^2 \cdot 1024^2 \cdot 4 = 17.2$\,GB at
$B{=}64$ and $128^2 \cdot 1024^2 \cdot 4 = 68.7$\,GB at $B{=}128$
(quadratic in $B$). The idealised
unified-vs-vanilla ratio has a closed form independent of $B$:
$(2 B^2 L_q L_d \cdot 4)/(B^2 L_q \cdot 4) = 2 L_d = 2048\times$.
The vanilla path peaks at \emph{at least} twice the
forward intermediate because $\nabla S$ has the same shape as $S$;
\flash{} replaces both with the saved argmax. Empirical step peaks
under the same matched-precision protocol (Tab.~\ref{tab:train},
$51.8$\,GB at $B{=}64$) sit $\sim\!1.5\times$ above this floor
because PyTorch's autograd materialises additional FP32 workspace
tensors during the multi-stage \texttt{einsum} backward beyond the
bare $S$ and $\nabla S$; the qualitative ordering (vanilla $\gg$
\flash{} CSR $\gg$ \flash{} unified) is preserved. The CSR-build
transients are the measured values of the shipped int32
construction. Under pure-FP16 in-dtype the vanilla peak drops to
$\sim\!16$\,GB at $B{=}64$ (Fig.~\ref{fig:training_parity}) since
both $S$ and $\nabla S$ are FP16 rather than FP32.
\end{table}

\begin{table}[h]\centering\small
\caption{Measured bare-step peak decomposition at the
Tab.~\ref{tab:bwd_mem} shapes (A100-80GB, \texttt{sum().backward()},
per-method \texttt{reset\_peak\_memory\_stats}). Every byte of the
measured peak has a named owner; the gap to the idealised floor of
Tab.~\ref{tab:bwd_mem} is inputs, autograd-allocated input gradients,
and workspace, which any scorer pays. Tab.~\ref{tab:train}'s slightly
higher absolute flash figures ($0.24$/$0.39$\,GB) additionally include
that harness's resident $100$\,MB L2-flush buffer and timing
scaffolding; vanilla agrees across the two protocols ($51.8$ vs.\
$51.66$\,GB). Source: \texttt{e4\_mem\_audit.json}.}
\label{tab:mem_audit}
\setlength{\tabcolsep}{3.5pt}
\begin{tabular}{l l r r r r r}
\toprule
$B$ & path & peak & $Q{+}D$ & $\nabla Q,\nabla D$ & argmax & transient \\
\midrule
$64$  & vanilla   & $51.66$ & $0.03$ & $0.03$ & --- & $51.59$ \\
      & unified   & $\mathbf{0.11}$ & $0.03$ & $0.03$ & $0.02$ & $0.03$ \\
      & CSR       & $0.28$ & $0.03$ & $0.03$ & $0.02$ & $0.20$ \\
$128$ & vanilla   & \textbf{OOM} & $0.07$ & $0.07$ & --- & --- \\
      & unified   & $\mathbf{0.25}$ & $0.07$ & $0.07$ & $0.07$ & $0.05$ \\
      & CSR       & $0.96$ & $0.07$ & $0.07$ & $0.07$ & $0.76$ \\
\bottomrule
\end{tabular}
\par\smallskip
\footnotesize All values decimal GB.
\end{table}

\section{Complete-Training-Step Memory and Wall-Clock}
\label{app:fullstep}
Tab.~\ref{tab:train} is operator-level by design (\S\ref{sec:eval},
``Why operator-level''). Here we report \emph{complete} training
steps: encoder activations, model parameters, optimizer state, and
weight gradients all included, with identical trainer code in both
columns so the scorer is the only variable. ``Ref'' is the reference
PyTorch scorer; ``ours'' is \flashshort{}. Reductions are computed
from unrounded measurements; these are observed complete-peak
differences, not a byte-level allocation split (components can peak
at different phases of the step).

\begin{table}[h]\centering\small
\caption{Complete-training-step peak memory (GB), A100-80GB.
ColBERT: BERT-base on MS~MARCO triplets, AdamW, bf16,
sentence-transformers trainer. ColPali: colpali-v1.2, LoRA $r{=}32$,
REAL-MM-RAG pages, gradient checkpointing, single-step memory probe.
$83.1$\,GB${}={}77.4$\,GiB; the A100 holds $80$\,GiB. Source:
\texttt{data/e1\_results/}.}
\label{tab:fullstep}
\setlength{\tabcolsep}{3pt}
\begin{tabular}{l r r r l}
\toprule
setting & $B$ & ours & ref & peak owner \\
\midrule
ColBERT, plain        & $64$   & $12.1$ & $12.2$ & encoder \\
ColBERT, plain        & $128$  & $22.3$ & $22.9$ & encoder \\
ColBERT, plain        & $256$  & $42.6$ & $45.2$ & scorer visible \\
ColBERT, plain        & $512$  & $\mathbf{83.1}$ & \textbf{OOM} & feasible only w/ ours \\
ColBERT, grad-ckpt    & $1024$ & $\mathbf{26.4}$ & $67.9$ & ref scorer ($-41.5$) \\
ColBERT, grad-ckpt    & $2048$ & $\mathbf{51.5}$ & \textbf{OOM} & ref block ${\approx}97$\,GB \\
ColPali, text query   & $128$  & $61.2$ & $61.2$ & encoding \\
ColPali, page-as-query& $96$   & $\mathbf{62.8}$ & $76.1$ & scorer contributes \\
\bottomrule
\end{tabular}
\end{table}

The ColPali text-query row shows no difference for a structural
reason: the peak forms while encoding (${\approx}0.5$\,GB per page
even with checkpointing) and the text-query similarity block
(${\approx}4$\,GB with its gradient) never rises above that
watermark. Within a fixed configuration, encoder-state memory grows
approximately linearly in $B$ while the similarity block grows as
$B^2 L_q L_d$, so a crossover always exists: small encoders cross
before the hardware limit (the ColBERT rows), and large encoders
cross once they adopt the cached-encoding
(GradCache-style~\cite{gradcache}) recipe,
where the block becomes the dominant large-batch term
(${\approx}27$\,GB${}+{}$gradient at $B{=}512$ with text queries,
versus \flashshort{}'s $54$\,MB argmax map).

\paragraph{Wall-clock.} Measured over matched equal-data training
runs (equal step counts per $B$, so ratios also apply to mean step
time): within $0.3\!-\!7.8\%$ at $B{=}64/128/256$ ($985$ vs $988$,
$840$ vs $873$, $778$ vs $843$\,s) where the encoder dominates, and
$\mathbf{1.26\times}$ faster end-to-end at $B{=}1024$ with both
sides checkpointed ($1{,}003$ vs $1{,}269$\,s).

\section{Does the Unlocked Batch Improve the Retriever?}
\label{app:batchquality}
We isolate the batch-size (negative-pool) effect with a controlled
protocol: equal examples per arm ($\text{max\_steps}=N/B$, so larger
batches do not see more data); learning rate scaled
$\sqrt{B/B_0}$ identically across arms, plus a fixed-LR control
($3$ seeds); $5$ seeds per point; evaluation on four NanoBEIR sets
fixed in advance (NFCorpus, SciFact, NQ, FiQA2018), mean nDCG@$10$,
with every arm scored through the \emph{reference} implementation so
quality never depends on our kernel. $B$ counts triplets per batch
($2B{-}1$ in-batch negatives per query); the retriever is a
BERT-base ColBERT trained on MS~MARCO (Tab.~\ref{tab:batchquality}).

\begin{table}[h]\centering\small
\caption{Equal-data batch-size$\to$quality curve ($5$ seeds per
point). The complete-step peak column is the ``ours'' column of
Tab.~\ref{tab:fullstep}. Source:
\texttt{data/e1\_results/b*\_s*.json}.}
\label{tab:batchquality}
\begin{tabular}{r c c r}
\toprule
$B$ & nDCG@$10$ mean & per-seed range & peak \\
\midrule
$64$  & $0.4264$ & $0.4230\!-\!0.4316$ & $12.1$\,GB \\
$128$ & $0.4370$ & $0.4188\!-\!0.4538$ & $22.3$\,GB \\
$256$ & $0.4441$ & $0.4300\!-\!0.4517$ & $42.6$\,GB \\
$512$ & $\mathbf{0.4611}$ & $0.4517\!-\!0.4673$ & $83.1$\,GB \\
\bottomrule
\end{tabular}
\end{table}

The mean improves monotonically: $+1.1$/$+1.8$/$+3.5$ points at
$B{=}128/256/512$ vs $B{=}64$ ($+8\%$ relative at the endpoint), and
every $B{=}512$ seed beats every $B{=}64$ seed --- the very batch the
complete-step table unlocks (the reference OOMs at $B{=}512$). We do
not claim a significant $B{=}128$ gain, since its seed range overlaps
$B{=}64$'s; the supported findings are the monotonic mean trend and
the non-overlapping $B{=}64$/$B{=}512$ seed ranges. The endpoint
separation also survives the fixed-LR control ($+2.4$ points,
$0.4387$ vs $0.4150$, $3$ seeds, again non-overlapping), so the
effect is not an LR-schedule artifact. Single-seed grad-checkpointing
probes at $B{=}1024/2048$ ($0.4663$/$0.4641$) land at the $B{=}512$
level, suggesting but not establishing saturation past
$B{\approx}512$ on this task. Reference and \flashshort{} runs match
within $\pm 0.5$ points at every matched pair, consistent with the
expected numerical equivalence of the exact scorer. This experiment
establishes the mechanism on a text retriever; we do not claim
measured quality gains for page-as-query ColPali.

\section{Comparison to Distributed Cross-Device Negatives}
\label{app:ddp}
Cross-device negatives enlarge the global negative pool, but each
device still scores its local queries against the gathered global
pool. The standard differentiable-all-gather formulation therefore
materialises a quadratic score block: with $W$ GPUs, per-device
memory scales as $O(B^2/W)$, so for a fixed budget the reachable
pool grows only as $\sqrt{W}$. Our \emph{streamed} sharded loss
computes the same InfoNCE cross-entropy objective over the global
pool (document sharding + short-side query communication + two-level
online softmax for the pool normaliser, whose backward recomputes
each \maxsim{} tile and its argmax) and keeps per-device memory
linear in $B$. A chunked/recompute
implementation can also shrink the resident block but still gathers
or replicates the long document side (the single-device recompute
trade-off is measured in Tab.~\ref{tab:train}). We measured at
$W{=}1,2,4$ (A100-80GB, \texttt{torchrun}+NCCL, ColBERT shapes
$L_q{=}32$/$L_d{=}180$, fp16, forward+backward, $5$-step means;
Tab.~\ref{tab:ddp}). Gradients are checked in-script against the reference on identical
data at every cluster size (cosine $\geq 0.9999998$ on $\nabla D$
and $\nabla Q$, rel.\ err.\ $\leq 3.0{\times}10^{-4}$).

\begin{table}[h]\centering\small
\caption{Distributed global-negative training: peak GB/device and,
where measured, step time (the conv.\ $W{=}2$ column reports peak
memory only). ``conv.'' = differentiable all-gather of documents
+ reference einsum on the global pool; ``ours'' = streamed sharded
loss (the \texttt{fm-flash-tiled} arm of the shipped results).
$B$ = global document pool. Source:
\texttt{e2\_ddp\_W\{1,2,4\}.json}.}
\label{tab:ddp}
\setlength{\tabcolsep}{2.5pt}
\begin{tabular}{r r r r r}
\toprule
pool $B$ & conv.\ $W{=}2$ & conv.\ $W{=}4$ & ours $W{=}1$ & ours $W{=}4$ \\
\midrule
$512$    & $9.21$ & $4.65$ / $30$\,ms & $0.22$ / $14$\,ms & $0.08$ / $7$\,ms \\
$1024$   & $36.5$ & $18.3$ / $102$\,ms & $0.39$ / $57$\,ms & $0.12$ / $17$\,ms \\
$2048$   & \textbf{OOM} & $72.8$ / $985$\,ms & $0.86$ / $230$\,ms & $0.25$ / $58$\,ms \\
$4096$   & --- & \textbf{OOM} & $1.49$ / $1.3$\,s & $0.64$ / $285$\,ms \\
$16{,}384$ & --- & --- & $4.84$ / $21$\,s & $3.69$ / $5.3$\,s \\
$65{,}536$ & --- & --- & $18.5$ / $333$\,s & $\mathbf{14.0}$ / $85$\,s \\
\bottomrule
\end{tabular}
\end{table}

Conventional memory follows the predicted law (it grows
${\approx}4\times$ when $B$ doubles and falls ${\approx}2\times$
from $W{=}2$ to $W{=}4$), and ours is $4.2\!-\!17\times$ faster at
matched $W{=}4$. Most directly: at $B{=}2{,}048$, ours on \emph{one}
GPU is $4.3\times$ faster than conventional on \emph{four}
($230$ vs $985$\,ms) at $0.86$\,GB versus $72.8$\,GB per
conventional rank. The $\geq\!16$K rows establish scaling capacity,
not practical operating points; step time there is dominated by the
$O(B^2)$ scoring FLOPs that any exact global-negative loss must pay.
Distribution and streaming compose: distribution parallelises
scoring, streaming removes the quadratic resident state.

\section{Serving-Path Measurements}
\label{app:pipeline}
\paragraph{Exact-reranking frontier at increasing $K$.} At visual
(ColPali) shapes ($L_d{=}1030$, $d{=}128$, fp16 candidates resident;
$K$ swept to $200$K, where the candidate embeddings alone reach
$53$\,GB on A100-80GB), the maximum feasible exact reranking depth
depends on the implementation. Unchunked eager OOMs at $K{=}64$K for
batched ($32$ concurrent text queries) or page-as-query serving;
deployed chunking reaches $K{=}200$K but pays an increasing latency
penalty; \flashshort{} reaches $K{=}200$K at $1.9\!-\!2.7\times$
lower latency than deployed chunking, with auxiliary storage of only $O(K)$ score/output
state (at unchunked's last surviving point, $K{=}32$K at $77$\,GB
peak, ours peaks at $8.5$\,GB --- the embeddings themselves).

\paragraph{Complete pipeline.} HotpotQA-500K with real ColBERTv2
embeddings and qrels (A100-80GB, single-stream): query encoding
$8.9$\,ms, exact single-vector candidate generation over all $500$K
documents $0.21$\,ms, candidate gather $0.09\!-\!5.7$\,ms, then the
dense rerank as the only variable. At $K{=}100$K the chunked
baseline costs $7.65$\,ms (comparable to the entire query encoding)
versus $2.7$\,ms for ours: a $28\%$ end-to-end throughput difference
($57.1$ vs $44.5$ QPS) from swapping nothing but the operator.
Retrieval quality rose monotonically with depth in this run
(nDCG@$10$ $0.2116\!\to\!0.2565$, $K{=}1$K$\to$$100$K), as deeper
exact reranking recovers first-stage misses. PLAID/ANN candidate
selection is orthogonal to the exact scorer and composes with it; the
exact first stage holds candidate selection fixed so the operator is
the only variable. Source: \texttt{e5\_pipeline.json},
\texttt{e6\_rerank\_frontier.json}.

\section{Related Work: Extended}\label{app:related_extended}
\paragraph{Flash-KMeans delta.}
\textbf{(i) Dual-asymmetry split.} \maxsim{}'s forward maps every
query token to exactly one document token, so $\nabla Q$ is a pure
gather (no scatter, no collisions) and only $\nabla D$ inherits the
contention; we therefore run $\nabla Q$ as a per-source kernel and
only build the CSR machinery for $\nabla D$. Flash-KMeans' two
centroid-side updates are symmetric and both go through the
sort-inverse path. \textbf{(ii) In-autograd CSR construction.}
Our \texttt{bincount}$\to$\texttt{cumsum}$\to$\texttt{argsort} runs
at each backward call inside the autograd Function from the
forward's saved argmax, with no precomputed assignment buffer carried
across iterations (Flash-KMeans precomputes per-iteration
assignments). \textbf{(iii) Fused unified backward.} A single-kernel
variant that hoists $Q$ rows into registers and writes $\nabla Q$
and $\nabla D$ together (\S\ref{sec:eval}, Tab.~\ref{tab:bwd_abl})
is, to our knowledge, not present in Flash-KMeans; it is the
dispatcher default for in-batch negatives because it matches CSR's
speed within $\sim\!8\%$ on the ColPali (in-batch) rows of this H100
ablation ($7\!-\!17\%$ on A100, Tab.~\ref{tab:train}; the small
ColBERT rows vary more) at $1.3$ to $4.7\times$ less peak memory.
The retrieval-side contributions — INT8$\times$INT8, padding-free
\texttt{cu\_seqlens}, and end-to-end nDCG parity at ColPali scale —
are out of scope for Flash-KMeans entirely.

\paragraph{Late-interaction serving stacks.}
ColBERTv2 \cite{colbertv2} and PLAID \cite{plaid} target a
complementary problem: offline residual compression of the document
representation plus centroid-pruned approximate scoring over a
compressed index across millions of documents. PLAID's wins come
from skipping work (only a small fraction of $(q,d)$ pairs are
scored) and from reducing per-pair byte count via residual codes;
the remaining exact dense \maxsim{} call on the surviving candidates
is still a dense $L_q\!\times\!L_d\!\times\!d$ reduction, and that
is the operator \flash{} accelerates. The two are stackable:
PLAID-style pruning chooses \emph{which} candidates to score,
\flash{} accelerates \emph{how} each surviving $(q,d)$ pair is
scored on the GPU, and is the natural back-end for the residual rerank
stage as well as for ColPali, where PLAID-style centroid/residual
machinery is not yet the standard deployment path and the dense
page-embedding footprint is much larger.

\paragraph{Compiler limits in detail.}
For \maxsim{}, Inductor cannot remove the materialised $[B,L_q,L_d]$
intermediate without rewriting the reduction structure: the einsum
produces a four-dimensional tensor whose layout the compiler treats
as the IR boundary, and the subsequent \texttt{max} and \texttt{sum}
are fused on \emph{that} tensor rather than into the matmul. As a
result \texttt{torch.compile(max-autotune)} narrows the dense-forward
gap (it closes part of the cuBLAS-vs-tensor-core inefficiency) but
does not change the asymptotic memory profile; all three OOM cliffs
in Tab.~\ref{tab:mem} reproduce under \texttt{compile-MA} in our
measurements. The fusion this paper does is a structural rewrite
(the running max replaces the materialised tensor entirely): a
transformation that current Triton-style operator authoring
expresses but Inductor's reduction lowering does not.

\section{Limitations: Extended}\label{app:limitations_extended}
\paragraph{Launch-bound regime.} At very small shapes
($L_q,L_d\!\lesssim\!64$, $B$ small) the kernel-launch overhead
dominates the per-call arithmetic, so \flash{} is at-parity rather
than faster, and the win is recovered only above the launch-bound
break-even.
\paragraph{Atomic-friendly backward.} When the expected number of
sources per destination is near one (sources sparse relative to the
destination set), atomic-scatter has near-zero contention and
can match or beat the CSR construction; the plain-atomic path is
selected for call signatures such as \texttt{shared\_docs=False} or
by explicit override (\S\ref{sec:backward}).
\paragraph{Top-$k$ ($k > 1$) reductions.} The backward saves a
single argmax index per source; a top-$k$ operator would need a
$[L_q,k]$ saved index buffer and a $k$-way CSR variant, which we do
not implement.
\paragraph{Tied maxima.} The running-max reduction is order-invariant
only for strict argmax; with engineered ties we follow PyTorch's
first-occurrence convention.
\paragraph{Open evaluation gaps.}
The single-step training speedup is modest at matched precision; the
dominant training win is memory and the batch-size unlock, not step
latency. INT8 ranking is validated on text distributions; the
REAL-MM-RAG quantised-path nDCG is reported only at a subset of
shapes (Tab.~\ref{tab:int8_realmmrag}).
The INT8 forward latency relative to \emph{FP16 \flash{}} is modest
($1.07$ to $1.27\times$ across the B-sweep at ColPali) because both
implementations are already streaming-light at $d{=}128$; the
larger INT8 wins ($3\!-\!6\times$) are over naive FP32 and
dequant-then-FP32-einsum baselines that some production rerankers
still run. We do not compare to ColBERT-PLAID-style compressed-index
scoring stacks (a different operating point: offline indexing +
approximate retrieval over millions of documents).
\paragraph{Hardware coverage.}
A100-SXM4 (40 and 80\,GB) and H100 80\,GB only. The kernel is
portable Triton and should compile on Volta/Turing or AMD MI, but
tile sizes were swept on A100/H100 and may need re-tuning.
\paragraph{Threats to validity.}
Forward and training tables are measured on synthetic tensors
matched to ColBERT / ColPali shapes; end-to-end ranking parity
against an FP32 reference is reported on two BEIR text sets
(Tab.~\ref{tab:beir_ndcg}) and on the four REAL-MM-RAG vision
subsets under two encoders (Tab.~\ref{tab:colpali_ndcg}). Because
\flash{} is an exact scorer replacement, ranking parity is sufficient
to guarantee identical downstream metrics for any fixed candidate
set; we do not run a full ViDoRe leaderboard-style pipeline (ANN /
PLAID-style pruning + scorer) end-to-end. Performance may vary with
GPU architecture, sequence-length distribution, and the launch-config
table; we report the heuristic-table choice but did not exhaustively
re-sweep on non-A100/H100 SKUs.

\section{Backward Pseudo-Code and Dispatch}\label{app:bwd_algo}

\begin{figure}[h]\centering\small
\begin{tikzpicture}[
  qtok/.style={circle, draw=blue!70, fill=blue!8, minimum size=14pt, inner sep=0pt, font=\scriptsize},
  dtok/.style={circle, draw=orange!85, fill=orange!12, minimum size=14pt, inner sep=0pt, font=\scriptsize},
  hot/.style={circle, draw=red!75, line width=0.8pt, fill=red!12, minimum size=15pt, inner sep=0pt, font=\scriptsize},
  arr/.style={->, >=Stealth, gray!70, thin},
  hotarr/.style={->, >=Stealth, red!75, semithick},
]
\node[qtok] (q0) {$q_0$};
\node[qtok, right=4mm of q0] (q1) {$q_1$};
\node[qtok, right=4mm of q1] (q2) {$q_2$};
\node[qtok, right=4mm of q2] (q3) {$q_3$};
\node[qtok, right=4mm of q3] (q4) {$q_4$};
\node[qtok, right=4mm of q4] (q5) {$q_5$};
\node[left=2mm of q0, font=\scriptsize\itshape] {sources $(i,s)$};

\node[dtok, below=12mm of q0, xshift=12mm] (d0) {$d_0$};
\node[dtok, right=10mm of d0] (d1) {$d_1$};
\node[hot,  right=10mm of d1] (d2) {$d_2$};
\node[dtok, right=10mm of d2] (d3) {$d_3$};
\node[left=2mm of d0, font=\scriptsize\itshape] {destinations $t$};

\draw[arr]    (q1) -- (d0);
\draw[arr]    (q2) -- (d1);
\draw[hotarr] (q0) -- (d2);
\draw[hotarr] (q3) -- (d2);
\draw[hotarr] (q4) -- (d2);
\draw[arr]    (q5) -- (d3);
\end{tikzpicture}
\caption{Backward gather/scatter asymmetry. Each source query token
$(i,s)$ picks exactly one destination document token $t^\star$ in
the forward (one arrow per $q$). $\nabla Q$ thus reads exactly one
$D$-row per program: embarrassingly parallel \emph{gather}, no
collisions. $\nabla D$ is the inverse: many sources may land on the
same destination ($d_2$ here, shown in red), giving data-dependent
\emph{scatter} contention that a naive \texttt{atomicAdd} would
serialise on. The inverse-grid CSR construction bins sources by
destination so each $\nabla D$ row is written once, atomic-free
(\S\ref{sec:backward}, Alg.~\ref{alg:bwd}).}
\label{fig:bwd_asym}
\end{figure}

\begin{algorithm}[h]
\caption{\flash{} backward (inverse-grid CSR), atomic-free}
\label{alg:bwd}
\small
\begin{algorithmic}[1]
\Require forward argmax $A \in \mathbb{Z}^{N_q \times B \times L_q}$
($A[q,b,i]$ = winning doc token for query token $(q,i)$ vs.\ doc $b$);
upstream grad $\nabla s \in \mathbb{R}^{N_q \times B}$;
$Q \in \mathbb{R}^{N_q \times L_q \times d}$, $D \in \mathbb{R}^{B \times L_d \times d}$
\Statex \textit{// 1. invert the source $\to$ destination map (once, in autograd)}
\State $\mathrm{dst}[q,b,i] \gets b\,L_d + A[q,b,i]$ \Comment{flat dest index into $[0,B L_d)$}
\State $\mathrm{row\_ptr} \gets \mathrm{concat}([0],\, \mathrm{cumsum}(\mathrm{bincount}(\mathrm{flatten}(\mathrm{dst}), \mathrm{minlength}{=}B L_d)))$ \Comment{$[B L_d{+}1]$ int32}
\State $\mathrm{col\_idx} \gets \mathrm{argsort}(\mathrm{flatten}(\mathrm{dst}))$ \Comment{source ids $(q,b,i)$ bucketed per dest}
\Statex \textit{// 2. $\nabla D$: one program per destination row, no two programs share a row}
\For{each destination row $r \in [0, B L_d)$ \textbf{in parallel}}
  \State $b \gets \lfloor r / L_d \rfloor$;\ \ $g \gets 0 \in \mathbb{R}^{d}$ \Comment{all sources in row $r$ share doc $b$; FP32 register}
  \For{source $(q,b,i) \in \mathrm{col\_idx}[\mathrm{row\_ptr}[r]:\mathrm{row\_ptr}[r{+}1]]$}
     \State $g \mathrel{+}{=}\ \nabla s[q,b]\cdot Q[q,i,:]$ \Comment{FP32 acc, FP16 operand}
  \EndFor
  \State $\nabla D[r] \gets g$ \Comment{single coalesced write, \textbf{no atomics}}
\EndFor
\Statex \textit{// 3. $\nabla Q$: one program per $(q,i)$, gathers winning $D$ rows}
\For{each $(q,i) \in [0,N_q) \times [0,L_q)$ \textbf{in parallel}}
  \State $\nabla Q[q,i,:] \gets \sum_{b}\, \nabla s[q,b]\, D[b,\, A[q,b,i],\, :]$ \Comment{$B$ gathers, no collision}
\EndFor
\end{algorithmic}
\end{algorithm}

\paragraph{Dispatch table.} Three backward variants ship; the
runtime dispatcher picks one at each \texttt{.backward()} call. The
criteria are deterministic (no autotune trial) and depend only on
the call signature, as in the table below (plain atomic scatter is
preferred only where the expected per-destination collision count is
near $1$, i.e.\ sources sparse relative to destinations):

\begin{center}\small
\begin{tabular}{@{}l l p{0.44\linewidth}@{}}
\toprule
Variant & Picked when & Why \\
\midrule
\emph{atomic\_unified} & \texttt{shared\_docs=True} (default) & fused $\nabla Q\!+\!\nabla D$, $1.3\!-\!4.7\times$ leaner than CSR; within $\sim\!8\%$ of CSR latency on H100, $7\!-\!17\%$ on A100 \\
\emph{atomic}          & \texttt{shared\_docs=False} (KD path) & per-query doc set; Q-hoisting does not apply, plain atomic scatter is the only correct option here \\
\emph{invgrid CSR}     & \texttt{FLASH\_BWD\_PATH=invgrid} (opt-in) & atomic-free destination-owned scatter (no FP32 atomics) \\
\bottomrule
\end{tabular}
\end{center}

No shape thresholds, no autotune dispatch: every call to the same
backward path takes the same code path. \texttt{FLASH\_BWD\_PATH}
also accepts \texttt{atomic} and \texttt{atomic\_unified} as explicit
overrides for the ablation in \S\ref{sec:eval}.

\section{INT8 Kernel: Deferred Dequantisation}\label{app:int8_algo}

\flash{}'s INT8 win over the production dequant-then-naive baseline
(\S\ref{sec:int8}) comes not from quantisation itself but from a
\emph{deferred-dequant} kernel pattern that keeps $D$ in INT8
through the entire memory hierarchy and applies the per-row scale
to the post-dot scalar instead of to each input element. The
algebraic trick uses linearity of the dot product over the per-token
scales, each constant along the embedding axis:
$\langle Q_i^{\text{deq}}, D_{j,t}^{\text{deq}}\rangle
 = q_{s,i}\, d_{s,j,t}\, \langle Q_i^{\text{int8}}, D_{j,t}^{\text{int8}}\rangle$,
where $q_{s,i}$ and $d_{s,j,t}$ are the per-token query and document
scales (each shared across the $d$ embedding components); if only $D$
is quantised this reduces to the single-scale form
$\langle Q_i, D_{j,t}^{\text{int8}}\rangle\, d_{s,j,t}$. With
per-token symmetric quantisation (no zero-point offset), the result
is one INT8 tensor-core matmul followed by a constant number of
floating-point scale multiplications per output cell: two for
INT8$\times$INT8, one when only $D$ is quantised
(Alg.~\ref{alg:int8}).

\begin{figure*}[h]\centering
\begin{minipage}[t]{0.48\linewidth}
\begin{algorithm}[H]
\caption{Standard (materialised) \maxsim{}: forward and training
backward. The two HBM-bound steps are marked.}
\label{alg:naive}
\small
\begin{algorithmic}[1]
\State $S \gets Q D^{\!\top}$ \Comment{$[B,L_q,L_d]$ \textbf{written to HBM}}
\State $m \gets \mathrm{rowmax}(S)$; \ $\mathrm{score} \gets \mathrm{rowsum}(m)$ \Comment{$S$ re-read}
\Statex \textit{// training backward}
\State $\nabla S \gets$ upstream grad routed to argmax \Comment{re-materialise}
\State $\nabla Q,\nabla D \gets \texttt{atomicAdd}$ over $\nabla S$ \Comment{hot-token \textbf{write contention}}
\end{algorithmic}
\end{algorithm}
\end{minipage}\hfill
\begin{minipage}[t]{0.48\linewidth}
\begin{algorithm}[H]
\caption{\flash{} INT8$\times$INT8 forward inner loop (one
$(Q_{\text{blk}}, D_{\text{blk}})$ tile-pair; per-token symmetric
quant; INT8 tensor cores).}
\label{alg:int8}
\small
\begin{algorithmic}[1]
\Require $Q_{\text{blk}}\in\mathbb{Z}_{8}^{b_q\times d}$, $D_{\text{blk}}\in\mathbb{Z}_{8}^{b_d\times d}$, per-row scales $q_s, d_s$, FP32 acc $m$
\State $S_{\text{int32}} \gets$ \Call{tl.dot}{$Q_{\text{blk}}, D_{\text{blk}}^{\!\top}$} \Comment{INT8 TC, $\Theta(b_q b_d d)$ INT8 ops}
\State $S \gets S_{\text{int32}} \cdot q_s[:,\!\text{None}] \cdot d_s[\text{None},\!:]$ \Comment{$\Theta(b_q b_d)$ scale multiplications — dequant done}
\State mask invalid $d$ positions to $-\infty$
\State $m \gets \max(m,\ \mathrm{rowmax}(S))$
\end{algorithmic}
\end{algorithm}
\end{minipage}
\end{figure*}

\paragraph{Cost vs.\ a naive dequant-then-naive baseline.} The
naive path first dequantises $D$ (and $Q$) element-wise, costing
$(L_d{+}L_q)\,d$ FP multiplies per (query, doc) pair, materialises
the FP $D$ tensor, then scores in FP. \flash{}'s INT8 path instead
keeps $D$ in INT8 through HBM and SRAM (one byte per element), runs
the dot on INT8 tensor cores at $2\times$ the FP16 throughput, and
applies the per-token scales \emph{once per dot-product output cell} (before the rowmax)
(line~2) rather than to every $D$ element. The dequantised $D$ tensor
never exists in HBM, SRAM, or registers; the win is bandwidth and
tensor-core throughput (not a lower asymptotic multiply count, since
the deferred scale is applied per output cell). This, together with
the halved HBM bytes for $D$, is what produces the
$\mathbf{3}$--$\mathbf{6}\times$ end-to-end speedup over the
dequant-then-naive baseline reported in
Tab.~\ref{tab:int8} and Tab.~\ref{tab:int8_B}.

\paragraph{Affine (uint8) variant.} For an affine quantisation
$D^{\text{deq}}_{j,t} \!=\! D^{\text{u8}}_{j,t}\,s_{j,t} + m_{j,t}$
(used by the \texttt{quantize\_int8} path), the algebraic
deferral still holds with one extra term:
$\langle Q_i, D^{\text{deq}}_{j,t}\rangle
 = \langle Q_i, D^{\text{u8}}_{j,t}\rangle\, s_{j,t}
 + (\textstyle\sum_k Q_{i,k})\, m_{j,t}$, where $\sum_k Q_{i,k}$ is
precomputed once per query token (negligible). The symmetric path
(default) drops the $m$ term, leaving pure scale multiplications;
the affine offset term is where a fused multiply-add applies.

\section{Training-Parity Curve}\label{app:training_parity}
\begin{figure}[h]\centering
\includegraphics[width=\textwidth]{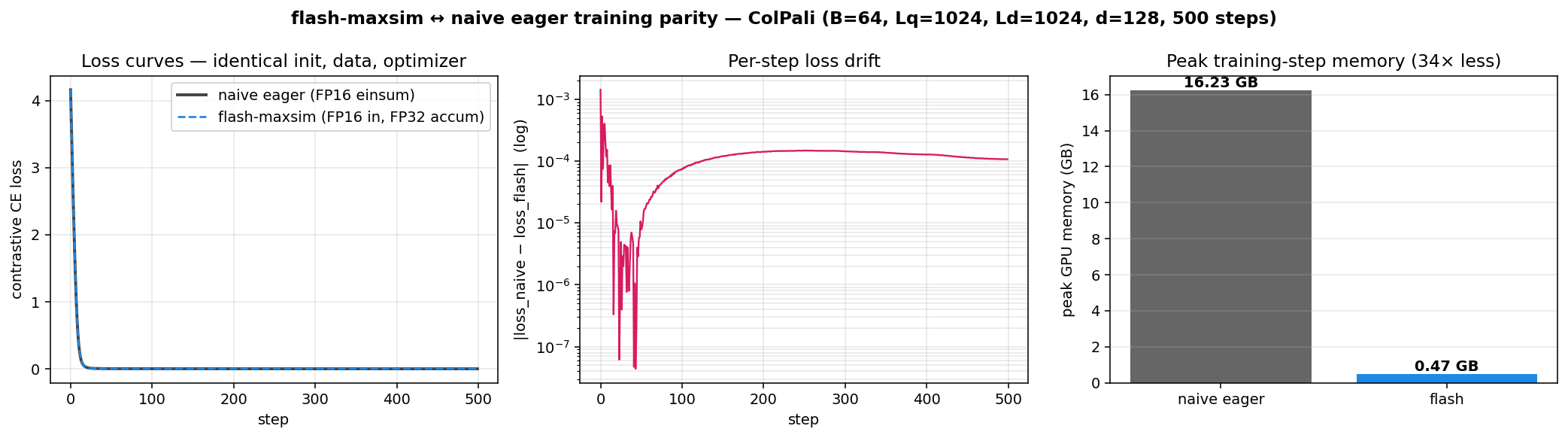}
\caption{End-to-end training parity at ColPali shape ($B{=}64$,
$L_q{=}L_d{=}1024$, $d{=}128$, $500$ contrastive steps, identical
seed and Adam state). The naive baseline here uses FP16 in-dtype
einsum reduction (no FP32 cast), distinct from the matched-FP32
protocol of Tab.~\ref{tab:train}; we use FP16-in-dtype here only to
let the naive path fit, since matched FP32 OOMs at $B{=}64$ on
$80$\,GB. \textbf{Left:} loss curves overlay
(visually indistinguishable). \textbf{Centre:} per-step absolute
drift $|\mathcal{L}_{\text{naive}}{-}\mathcal{L}_{\flash{}}|$ on
$\log$ scale, bounded by $1.4{\times}10^{-3}$, mean
$1.1{\times}10^{-4}$. \textbf{Right:} peak training-step GPU memory,
$34\times$ less for \flash{} because the $[B,B,L_q,L_d]$ FP16
similarity matrix and its gradient never form. At $B{=}128$ even
the FP16-in-dtype naive path OOMs on $80$\,GB at step 1; the parity
figure is therefore measured at the largest $B$ at which it still
fits.}
\label{fig:training_parity}
\end{figure}

\section{BEIR at Corpus Scale}\label{app:beir_scale}

\begin{table}[h]\centering\small
\caption{End-to-end retrieval parity on BEIR (ColBERTv2 encoder,
A100-80GB; ArguAna $Q{=}1406, N{=}8674$; SciDocs
$Q{=}1000, N{=}25657$). Each cell is the value shared by the FP32
\texttt{einsum} baseline and \flash{}; they match to four decimal
places. Per-query mean $|\Delta|\!\sim\!2{\times}10^{-4}$. SciDocs
omits Recall (binary qrels).}
\label{tab:beir_ndcg}
\begin{tabular}{llcccc}
\toprule
Dataset & $K$  & nDCG   & Recall & top-$K$         & top-$10$ exact \\
\midrule
ArguAna & $5$  & $.2844$ & $.5277$ & $\mathbf{100\%}$ & \multirow{2}{*}{$1405/1406$} \\
        & $10$ & $.3308$ & $.6707$ & $\mathbf{100\%}$ & \\
SciDocs & $5$  & $.1341$ & ---     & $\mathbf{100\%}$ & \multirow{2}{*}{$\mathbf{1000/1000}$} \\
        & $10$ & $.1565$ & ---     & $\mathbf{100\%}$ & \\
\bottomrule
\end{tabular}
\end{table}

Tab.~\ref{tab:beir_scale} extends the BEIR parity study
(Tab.~\ref{tab:beir_ndcg}) from the two small subsets to two
corpus-scale ones with precomputed ColBERTv2 embeddings,
$500$ queries each. HotpotQA ($500$K docs) runs both backends:
\flash{} (FP16 in / FP32 accum, streamed) and the chunked FP32
\texttt{einsum} reference; every retrieval metric is
\emph{identical} between them (parity max
$|\Delta|\!=\!6.4\!\times\!10^{-4}$ on raw scores, which never
crosses a rank boundary at the measured cells), and \flash{}
completes the $500{\times}500$K sweep $4.0\times$ faster
($26$\,s vs.\ $104$\,s wall). NQ ($2.68$M docs, $116$\,GB of FP16
embeddings, $1.4\times$ total VRAM) is scored out-of-core
(\S\ref{sec:ooc}); no GPU-resident dense baseline runs at this
size, so the row reports \flash{} alone: the point is that the
feasibility unlock produces a \emph{correct, complete} ranking
over $2.68$M documents, not a quality delta vs.\ an impossible
reference.

\begin{table}[h]\centering\small
\caption{Corpus-scale BEIR parity and feasibility (ColBERTv2 encoder, $500$
queries/subset, A100-80GB). HotpotQA: \flash{} and the chunked
FP32 reference produce identical metrics. NQ exceeds VRAM
($116$\,GB embeddings); \flash{} scores it out-of-core, and the dense
reference cannot run (marked ---).}
\label{tab:beir_scale}
\setlength{\tabcolsep}{4pt}
\begin{tabular}{lrcccc}
\toprule
Dataset & $B$ & backend & nDCG@$10$ & Recall@$10$ & wall \\
\midrule
HotpotQA & $500$K  & FP32 ref  & $.5681$ & $.6957$ & $104$\,s \\
         &         & \flash{}  & $.5681$ & $.6957$ & $26$\,s  \\
NQ       & $2.68$M & FP32 ref  & ---     & ---     & ---      \\
         &         & \flash{} (OOC) & $.5223$ & $.7460$ & $103$\,s \\
\bottomrule
\end{tabular}
\end{table}

\section{REAL-MM-RAG: Per-Subset Breakdown}\label{app:colpali_ndcg_full}

\begin{table}[h]\centering\small
\caption{REAL-MM-RAG end-to-end parity across four subsets
(FinReport, FinSlides, TechReport, TechSlides; $B{=}1674\!-\!2687$
pages each; ColPali $Q{=}100$/subset, GVE $Q{=}853\!-\!1354$/subset;
A100-80GB). nDCG@10 and Recall@10 shown as min--max across subsets;
ColPali baseline and \flash{} match to four decimal places at every
cell; GVE matches within $5{\times}10^{-4}$ on nDCG@$10$ and
$1.2{\times}10^{-3}$ on Recall@$10$; per-subset breakdown in
Tab.~\ref{tab:colpali_ndcg_full} below.}
\label{tab:colpali_ndcg}
\setlength{\tabcolsep}{4pt}
\begin{tabular}{lcccc}
\toprule
Encoder & nDCG@10        & Recall@10      & $\rho$           & top-10              \\
\midrule
ColPali & $.602\!-\!.866$ & $.760\!-\!.930$ & $\mathbf{1.0000}$ & $\mathbf{100\%}$    \\
GVE     & $.756\!-\!.932$ & $.947\!-\!.990$ & $0.9999$         & $\geq\!99.89\%$      \\
\bottomrule
\end{tabular}
\end{table}

\begin{table}[h]\centering\small
\caption{Per-subset breakdown of Tab.~\ref{tab:colpali_ndcg} (summary
above). $B$ pages per subset, single A100-80GB. \flash{}
runs FP16 inputs / FP32 accumulation; baseline is a chunked naive
einsum reference in FP32 throughout. Identical metrics to four
decimal places on ColPali; within $\leq 5\!\times\!10^{-4}$ on GVE
nDCG@$10$ ($\leq 1.2\!\times\!10^{-3}$ Recall@$10$),
where longer queries ($L_q\!\in\![88,112]$) compound a longer
sequence of FP16 products accumulated in FP32.}
\label{tab:colpali_ndcg_full}
\setlength{\tabcolsep}{3pt}
\resizebox{\columnwidth}{!}{%
\begin{tabular}{llrccccccc}
\toprule
& & & \multicolumn{2}{c}{nDCG@$10$} & \multicolumn{2}{c}{Recall@$10$} & & \\
\cmidrule(lr){4-5} \cmidrule(lr){6-7}
Encoder & Subset & $B$ & naive & \flash{} & naive & \flash{} & $\rho$ & top-$10$ overlap \\
\midrule
\multirow{4}{*}{ColPali v1.2}
& FinReport  & $2687$ & $0.6019$ & $\mathbf{0.6019}$ & $0.7600$ & $\mathbf{0.7600}$ & $\mathbf{1.000000}$ & $\mathbf{100\%}$ \\
& FinSlides  & $2280$ & $0.6385$ & $\mathbf{0.6385}$ & $0.7900$ & $\mathbf{0.7900}$ & $\mathbf{1.000000}$ & $\mathbf{100\%}$ \\
& TechReport & $1674$ & $0.8209$ & $\mathbf{0.8209}$ & $0.9300$ & $\mathbf{0.9300}$ & $\mathbf{1.000000}$ & $\mathbf{100\%}$ \\
& TechSlides & $1963$ & $0.8659$ & $\mathbf{0.8659}$ & $0.9300$ & $\mathbf{0.9300}$ & $\mathbf{1.000000}$ & $\mathbf{100\%}$ \\
\midrule
\multirow{4}{*}{GVE}
& FinReport  & $2687$ & $0.7562$ & $0.7566$ & $0.9472$ & $0.9484$ & $0.999999$ & $99.89\%$ \\
& FinSlides  & $2280$ & $0.8054$ & $0.8049$ & $0.9762$ & $0.9762$ & $0.999999$ & $99.95\%$ \\
& TechReport & $1674$ & $0.8764$ & $0.8765$ & $0.9784$ & $0.9784$ & $0.999999$ & $99.91\%$ \\
& TechSlides & $1963$ & $0.9321$ & $0.9321$ & $0.9904$ & $0.9904$ & $0.999999$ & $99.93\%$ \\
\bottomrule
\end{tabular}}
\end{table}

\begin{table}[h]\centering\small
\caption{INT8 \flash{} on REAL-MM-RAG, per subset (same shapes /
qrels as Tab.~\ref{tab:colpali_ndcg_full}). Each cell is
nDCG@$10$ for the approximate INT8 path; $\Delta$ is the signed
gap to the FP32 reference (negative means INT8 worse). Maximum
$|\Delta|$ across all 8 cells is $0.003$ at ColPali/FinReport;
INT8 \emph{matches or improves} the FP32 nDCG on $5$ of $8$ subsets
and is within $4{\times}10^{-4}$ on the other three.}
\label{tab:int8_realmmrag}
\begin{tabular}{llrcc}
\toprule
Encoder & Subset & $B$ & INT8 nDCG@$10$ & $\Delta$ vs FP32 \\
\midrule
\multirow{4}{*}{ColPali v1.2}
& FinReport  & $2687$ & $0.6048$ & $+0.0029$ \\
& FinSlides  & $2280$ & $0.6380$ & $-0.0004$ \\
& TechReport & $1674$ & $0.8209$ & $\phantom{+}0.0000$ \\
& TechSlides & $1963$ & $0.8659$ & $\phantom{+}0.0000$ \\
\midrule
\multirow{4}{*}{GVE}
& FinReport  & $2687$ & $0.7575$ & $+0.0013$ \\
& FinSlides  & $2280$ & $0.8051$ & $-0.0003$ \\
& TechReport & $1674$ & $0.8766$ & $+0.0002$ \\
& TechSlides & $1963$ & $0.9317$ & $-0.0004$ \\
\bottomrule
\end{tabular}
\end{table}

\noindent On \emph{text} BEIR, INT8 is even tighter than on the
vision subsets above: within $|\Delta|\!\leq\!4{\times}10^{-4}$ of the
FP32 reference at nDCG@$5$/$10$/$20$ (ArguAna $1406$ queries, SciDocs
$1000$), with top-$20$ overlap $\geq\!99.7\%$. The operator-level
long-doc top-$20$ drift (Tab.~\ref{tab:int8}) does not materialise as
end-to-end quality loss on either modality.

\section{Background: Extended}\label{app:background_extended}

\paragraph{Naive implementation pseudo-code.} The standard PyTorch
\maxsim{} forward + training backward with the two HBM-bound steps
marked is shown alongside the INT8 fused kernel in
Alg.~\ref{alg:naive}, App.~\ref{app:int8_algo}.

\begin{figure}[h]\centering
\includegraphics[width=0.6\textwidth]{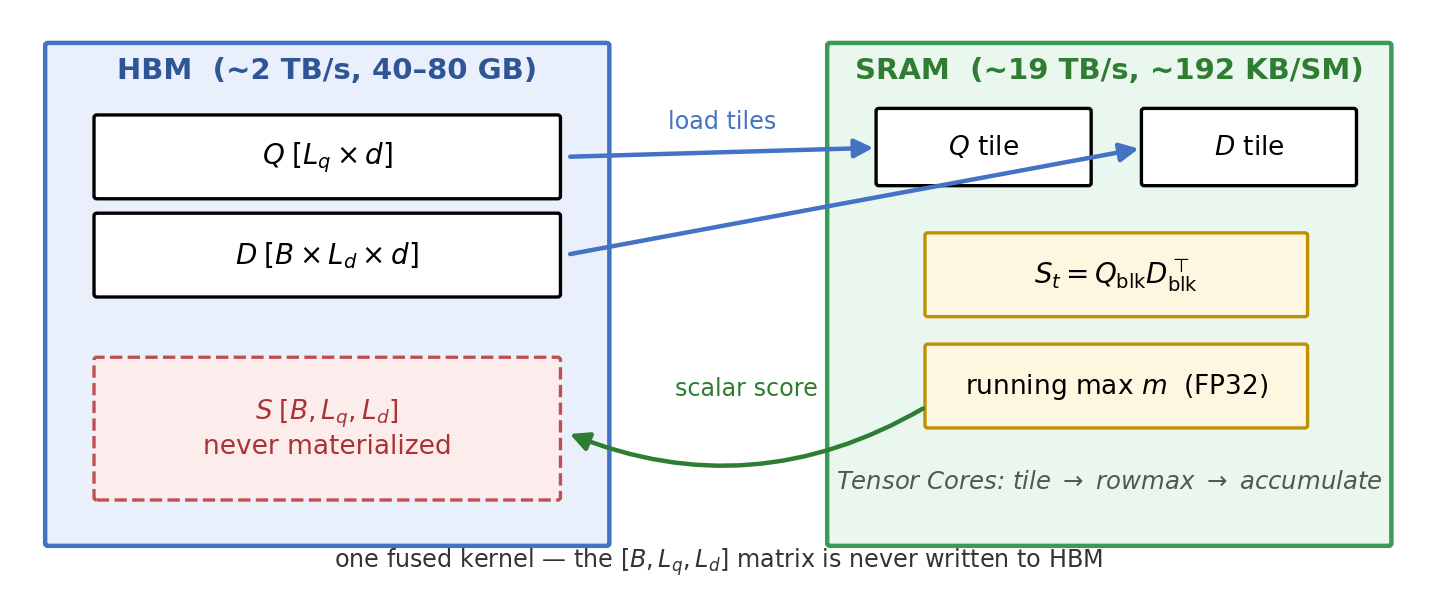}
\caption{\flash{} streams $Q$ and $D$ tiles from HBM into SRAM,
forms the sub-tile $S_t$ and folds it into a running max on chip,
and writes back only the scalar score. The $[B,L_q,L_d]$ similarity
tensor is never written to HBM.}
\label{fig:hierarchy}
\end{figure}

\paragraph{Memory hierarchy diagram.}
\flash{} streams $Q$ and $D$ tiles from HBM into SRAM, forms the
sub-tile $S_t$ and folds it into a running max on chip, and writes
back only the scalar score; the $[B, L_q, L_d]$ similarity tensor is
never written to HBM (Fig.~\ref{fig:hierarchy}).

\paragraph{Proof sketch and arithmetic intensity
(Thm.~\ref{thm:io}).} Each program of Alg.~\ref{alg:fwd} streams
its $Q$ and $D_j$ through SRAM one
$(Q_{\mathrm{blk}}, D_{\mathrm{blk}})$ tile-pair at a time, forms
$S_t$ on chip, and folds it into the running max; $S$ is never
written to or read from HBM, so operand traffic is
$\Theta(N_q B (L_q{+}L_d)d)$ and output traffic $\Theta(N_q B)$.
The naive path additionally writes and reads $S$, adding
$\Theta(N_q B L_q L_d)$, which dominates whenever
$L_q L_d \gg (L_q{+}L_d)d$. With only the operand reads as HBM
traffic, the fused arithmetic intensity becomes
$2 L_q L_d d / (2(L_q{+}L_d)d) = L_q L_d/(L_q{+}L_d) \approx 512$
FLOPs/byte at ColPali shape, above the H100 ridge: the matmul moves
from memory-bound to compute-bound: it is faster not because the
arithmetic is cheaper, but because it no longer waits on memory.

\paragraph{System-level constraints in real deployments.}
The materialised tensor imposes hard system limits, not just
slowdowns. At ColPali $B{=}10$K the similarity tensor alone occupies
$21$\,GB in FP16 ($42$\,GB in FP32), with total measured naive peaks
of $23.9$ and $47.2$\,GB respectively (Tab.~\ref{tab:mem}); by
$B{=}20$K both naive paths OOM, so reranking a
large candidate set is impossible without manual chunking. In
\emph{training}, in-batch-negatives scoring forms an all-pairs
$[N_q, B, L_q, L_d]$ tensor whose size is \emph{quadratic} in the
batch $B$; with autograd also retaining its gradient, the
contrastive batch size is capped far below what the model and
optimiser alone would allow; at ColPali shape, $B{=}128$ OOMs an
$80$\,GB GPU (\S\ref{sec:eval}). These caps, on corpus size at
inference and batch size at training, are the real-deployment cost
of materialisation.

\section{Method: Extended Discussion}\label{app:method_extended}

\paragraph{FP32-cosine-equivalent caveat.} The CSR-backward gradient
matches an FP32 reference to $\nabla Q / \nabla D$ cosine $1.0000$;
we use the phrase \emph{FP32-cosine-equivalent}: \emph{bit equality}
is neither expected nor claimed; the FP32-accumulated CSR scatter is
atomic-free (accumulation follows the built source list) but the
inner products are not bit-identical to a
different reduction order.

\paragraph{bf16 silent-failure mode (full analysis).}
The FP32-accumulation discipline of Tab.~\ref{tab:precision} matters
most for bf16 (\texttt{torch.bfloat16}) workloads, which are common
in modern training. The \emph{naive eager (bf16, in-dtype)} row of
Tab.~\ref{tab:precision} accumulates the $\max$- and sum-reductions
in bf16; at ColPali shape this sums thousands of products in bf16
and loses precision catastrophically: against a true-FP32 reference,
naive bf16 \maxsim{} at $L_q{=}L_d{=}1024$ shows max
$|\Delta|\!\approx\!0.5$ to $0.76$, large enough to flip top-$20$
rankings on ranking-tie-dense workloads. The corresponding \flash{}
(BF16) row of Tab.~\ref{tab:precision} promotes to FP32 inside the
kernel and so achieves max $|\Delta|\!\approx\!0.009$ to $0.014$,
i.e.\ \textbf{$35\times$ to $87\times$ tighter than the naive bf16
path} at the same shape. For production training and serving in
bf16, the choice is not between two equally-correct backends: on all
tested non-quantised workloads \flash{} preserves rankings
(top-$20$/$50$ overlap $100\%$ vs.\ the FP32 reference), while the
naive bf16 path does not.

\paragraph{Backward-path ablation: conservative-contention caveat.}
The H100 ablation (Tab.~\ref{tab:bwd_abl}) uses random unit-norm
embeddings, which produce a roughly uniform argmax distribution and
therefore a \emph{conservative} lower bound on atomic contention: on
real correlated embeddings the same $(b, t)$ destination is the
argmax for many more $(q, s)$ sources, so the CSR / unified backward
advantage over plain atomic scatter should grow, not shrink, in
production workloads.

\paragraph{CSR-path cost decomposition (full).}
To localise where the CSR backward spends its time (useful for
understanding why the fused unified backward, which has no CSR build
cost, matches CSR at large shapes) we instrument the CSR path on
the same shapes, for a single query ($N_q{=}1$), on A100-80GB and
time each component with CUDA events (Tab.~\ref{tab:bwd_breakdown}). The CSR build is the largest
single component ($54$ to $78\%$ of the backward) while the two
kernel passes ($\nabla Q$ and $\nabla D$) together are $20$ to
$40\%$. The CSR build has $\Theta(N_q B L_q + B L_d)$ work (linear in $B$
at fixed sequence lengths); its measured wall-clock grows
sub-linearly over the tested range ($0.45\!\to\!0.55$\,ms for
$B\,128\!\to\!1024$ in the ColBERT rows, from rising utilisation),
which is why CSR remains a defensible option even
at ColPali $B{=}128$. The forward kernel cost is shown for context;
the backward pieces (CSR build $+ \nabla Q + \nabla D$) and the
autograd-roundtrip measurement agree at the small ColBERT shapes but
diverge at ColPali (the roundtrip runs up to ${\sim}2\times$ higher),
where the autograd backward also performs a full FP32 $\nabla D$
allocation/cast not charged to the per-kernel pieces.

\end{document}